%% file: Main.tex
\documentclass[journal]{IEEEtran}
\usepackage{amsmath,amsthm,amssymb}
\usepackage{mathrsfs,mathtools,mathabx}
\usepackage{bm}
\usepackage{graphicx}
\usepackage{textcomp}
\usepackage{xcolor}
\usepackage{dsfont}
\usepackage{pgfplots}
\pgfplotsset{compat=newest}
\usepackage{subfigure}
\usepackage{cite}
\usepackage{booktabs}
\usepackage{float}
\usepackage{multirow}
\usepackage{xparse}
\usepackage{hyperref}
\usepackage{tikz}
\usetikzlibrary{calc}
\usetikzlibrary{spy}
\usetikzlibrary{arrows,decorations.pathreplacing,patterns}

\usepackage{etoolbox}

\graphicspath{{./Figures/}}

\newtheorem{prop}{Proposition}
	
\newtheorem{rem}{Remark}	
\newtheorem{lem}{Lemma}

\theoremstyle{definition}
\newtheorem{definition}{Definition}

\allowdisplaybreaks % Break equations into different lines
\usepackage{balance} % Balance columns last page

\definecolor{eggplant}{rgb}{0.38, 0.25, 0.32}
\definecolor{pearl}{rgb}{0.94, 0.92, 0.84}
\definecolor{chestnut}{rgb}{0.8, 0.36, 0.36}
\definecolor{airforceblue}{rgb}{0.36, 0.54, 0.66}
\definecolor{cadmiumorange}{rgb}{0.93, 0.53, 0.18}
\definecolor{bleudefrance}{rgb}{0.19, 0.55, 0.91}
\definecolor{carolinablue}{rgb}{0.6, 0.73, 0.89}
\definecolor{blue(ncs)}{rgb}{0.0, 0.53, 0.74}
\definecolor{dodgerblue}{rgb}{0.12, 0.56, 1.0}
\definecolor{cssgreen}{rgb}{0.0, 0.5, 0.0}
\definecolor{cadmiumgreen}{rgb}{0.0, 0.42, 0.24}
\definecolor{cadmiumorange}{rgb}{0.93, 0.53, 0.18}
\definecolor{amaranth}{rgb}{0.9, 0.17, 0.31}
\definecolor{bluegray}{rgb}{0.4, 0.6, 0.8}
\definecolor{cadmiumgreen}{rgb}{0.0, 0.42, 0.24}
\definecolor{amaranth}{rgb}{0.9, 0.17, 0.31}
\definecolor{amethyst}{rgb}{0.6, 0.4, 0.8}
\definecolor{amber}{rgb}{1.0, 0.75, 0.0}
\definecolor{azure}{rgb}{0.0, 0.5, 1.0}
\definecolor{babyblue}{rgb}{0.54, 0.81, 0.94}
\definecolor{bazaar}{rgb}{0.6, 0.47, 0.48}
\definecolor{celestialblue}{rgb}{0.29, 0.59, 0.82}
\definecolor{darklavender}{rgb}{0.45, 0.31, 0.59}
\definecolor{bluebell}{rgb}{0.64, 0.64, 0.82}
\definecolor{chamoisee}{rgb}{0.63, 0.47, 0.35}
\definecolor{darkcerulean}{rgb}{0.03, 0.27, 0.49}
\definecolor{iris}{rgb}{0.35, 0.31, 0.81}
\definecolor{jazzberryjam}{rgb}{0.65, 0.04, 0.37}
\definecolor{brown(web)}{rgb}{0.65, 0.16, 0.16}
\definecolor{palecgray}{rgb}{0.93, 0.94, 0.94}

\definecolor{egyptianblue}{rgb}{0.06, 0.2, 0.65}
\definecolor{burntumber}{rgb}{0.54, 0.2, 0.14}
	\definecolor{charcoal}{rgb}{0.21, 0.27, 0.31}
\definecolor{camel}{rgb}{0.76, 0.6, 0.42}
\definecolor{chamoisee}{rgb}{0.63, 0.47, 0.35}
	\definecolor{darkcerulean}{rgb}{0.03, 0.27, 0.49}

\begin{document}

\title{On Designing Modulation for Over-the-Air Computation —  Part II: Pyramid Sampling }

\author{ Saeed Razavikia,~\IEEEmembership{Member,~IEEE}, Carlo Fischione,~\IEEEmembership{Fellow,~IEEE}
\thanks{S. Razavikia and C. Fischione are with the School of Electrical Engineering and Computer Science, KTH Royal Institute of Technology, Stockholm, Sweden (e-mail: \{sraz, carlofi\}@kth.se). C. Fischione is also with Digital Futures of KTH. 

}
\thanks{S. Razavikia was jointly supported by the Wallenberg AI, Autonomous Systems and Software Program (WASP) and the Ericsson Research Foundation. The  SSF SAICOM project, the Digital Futures project DEMOCRITUS, and the Swedish Research Council Project MALEN partially supported this work.}
}

\maketitle

\begin{abstract}
Over-the-air computation (OAC) harnesses the natural superposition of wireless signals to compute aggregate functions during transmission, thereby collapsing communication and computation into a single step and significantly reducing latency and resource usage. In Part~I, digital OAC was formulated as a noise-aware constellation design problem by casting encoder design as a max–min optimization that aligns minimum Euclidean distances between superimposed constellation points with squared differences of their corresponding function outputs. 

In this paper, Part II, we address the prohibitive complexity and quantization challenges inherent in digital OAC constellation design for large-scale edge networks. More precisely,  we introduce a pyramid sampling strategy that judiciously selects a subset of superimposed constellation points to reduce the encoder design complexity from $\mathcal{O}(q^K)$ to $\mathcal{O}(q^{K-p+1})$, where $p\in\{1,\dots, K\}$ denotes the sampling order, $q$ levels of modulation, and $K$ denotes the number nodes in the network. Under the assumption of symmetric aggregation, this approach enables a controlled trade-off between computational complexity and function computation accuracy. As a special case, we propose majority-based sampling ($p=K$), which confines aggregation to only $q$ consensus points, inherently avoiding destructive overlaps and permitting the use of standard digital modulations (e.g., QAM, PSK, ASK) without bespoke constellation designs. We also show via several simulations, across various aggregation functions, modulation levels, and noise levels, that moderate sampling orders attain acceptable performance with orders-of-magnitude fewer constraints than exhaustive designs.    
\end{abstract}


%===================================================
\section{Introduction}
%===================================================

Edge‐intelligent applications—ranging from federated learning \cite{mcmahan2017communication,amiri2020federated,razavikia2024FedComp} and distributed sensor fusion~\cite{Golden2013Harnessing} and in IoT networks to real‐time analytics for autonomous systems\cite{chen2018over}—are placing unprecedented strain on wireless uplink capacity and introducing severe latency bottlenecks. For instance, aggregating gradient updates from hundreds of devices in federated learning can saturate the multiple‐access channel (MAC), degrading convergence speed and limiting scalability~\cite{amiri2020federated,hellstrom2022wireless}. Over‐the‐air computation (OAC) addresses these challenges by exploiting the natural superposition of signals in a wireless multiple‐access channel to compute aggregate functions directly during transmission, thereby integrating communication and computation into a single operation, and reducing both resource consumption and latency as the network scales~\cite{csahin2023survey,per2024waveforms}.

In Part~I~\cite{razavi2025revisit}, we cast digital OAC as a noise‐aware constellation design problem. Instead of targeting bit‐error minimization, we formulated a max–min optimization that aligns the minimum Euclidean distance between superimposed constellation points with the squared difference of their corresponding function outputs. More specifically, to design the constellation diagram of the transmitted signal by nodes, we revisit the optimization problem proposed in \cite{saeed2023ChannelComp} as follows: 
%------------
\begin{align}
 \label{eq:main_problem}
\max_{\bm{x},\lambda}~\lambda~~~\text{s.t.}~~\mathcal{D}_{\mathbb{C}}(r_i,r_j)\ge\lambda\bigl|f^{(i)}-f^{(j)}\bigr|^2,~~
  \|\bm{x}\|_2^2\leq 1,
\end{align}
%------------
for all $(i,j)$, where \(\bm{x}:=[x^{(1)},\ldots,x^{(q)}]^{\mathsf{T}}\in \mathbb{C}^q\) collects the \(q\) constellation points, i.e.,  and \(f^{(i)}\) denotes the function value associated with the noise-free received point \(r^{(i)}\). Then, we tailored the distance among the induced constellation points $\mathcal{D}_{\mathbb{C}}(r_i,r_j)$  to the distribution of the noise, which yields robust constellation diagrams for Gaussian, Laplace, and heavy‐tailed noise distributions. Indeed, we show that avoiding overlaps proposed by \cite{saeed2023ChannelComp} can be approximately translated as maximizing the minimum error among the constellation points over a noisy MAC.

However, on the one hand, as network scale $K\gg 1$ or modulation level $q$ grows, constellation optimization over possible $\mathcal{O}(q^K)$ pairs of output value becomes computationally prohibitive.  On the other hand, after the aggregation steps in any modern computation system, we require a sampling step due to the limited available memory at the server. This sampling or quantization step is unavoidable~\cite{higham2002accuracy}. These dual challenges compel a revisit of the optimization problem in \eqref{eq:main_problem} to identify a far more tractable approach. In Part II, we examine the complexity of solving the optimization problem given the modulation design in \eqref{eq:main_problem} and propose a sampling strategy for sampling the constellation points to reduce the optimization complexity while preserving computational accuracy.

\subsection{Why do We Need Sampling for Aggregation?}
IEEE 754--2019 specifies each floating-point format by a fixed precision \( b \), corresponding to the number of bits in the significand (including the implicit leading bit in normalized numbers), which yields \( q = 2^b \) distinct significand patterns per exponent value. Each real number \( x \in \mathbb{R} \) is mapped to the nearest representable floating-point value \( \hat{x} \in \mathcal{F} \) via rounding, typically under the round-to-nearest mode, introducing a maximum error of one-half unit in the last place  \cite{ISO/IEC,higham2002accuracy}.  If one were to evaluate a many‐to‐one mapping $f(x_1, x_2, \dots, x_K)$ over its entire domain, there would be up to \(q^K\) possible input combinations—for example, when \(K = 100\) and \(b = 16\), \((2^{16})^{100} \approx 10^{481}\). However, such cardinality makes exhaustive computation and unquantized storage infeasible. Such cardinality is not needed in practice, because the output of the function must be quantized as well to the same \(q\)-level grid, so that only \(\mathcal{O}(q)\) distinct function outputs remain. Although this might entail severe precision loss, the extremely large value of \(q\) ensures that the rounding error remains negligible in most practical scenarios. These principles underlie every digital computation: limited precision mandates quantization, and sampling is essential to constrain complexity, bound error, and enable feasible aggregation in large-scale computer networks and edge computing~\cite{chen2021quantization}.

% -----------
\input{Figs/Fig_SystemModel}
% -----------

Recent digital OAC schemes leverage sampling and quantization to manage both complexity and precision. The one-bit broadband digital aggregation (OBDA) quantizes each local update to its sign (two‐level sampling) and aggregates via majority voting, thereby bounding quantization error to one bit per device~\cite{zhu2020one}. Majority‐vote frequency-shift keying quantizes received energy on orthogonal tones, sacrificing phase for robust, non‐coherent detection at the cost of one tone per level~\cite{csahin2023distributed}. OFDM‐OBDA samples only symbol magnitudes across subcarriers, simplifying synchronization yet yielding coarse quantization that degrades low signal-to-noise ratio (SNR) performance~\cite{you2023broadband}. Huffman‐polynomial nulling embeds binary votes as polynomial zeros and samples nonzero coefficients, avoiding destructive overlaps but introducing polynomial‐evaluation errors~\cite{sahin2024over}. Type‐based multiple access allocates an orthogonal resource to each quantized output, sampling histogram counts for unbiased averaging but incurring high bandwidth per level~\cite{mergen2006type,per2024waveforms}. Balanced and radix numeral systems quantize digits in alternate bases to bound summation error, at the expense of larger alphabets and greater bandwidth~\cite{csahin2022over,tang2022radix}. Vector quantization in federated learning maps high‐dimensional gradients to shared codebook centroids, controlling distortion via codebook resolution but assuming device counts are much smaller than codebook size~\cite{qiao2024massive}. Finally, in \cite{saeed2023ChannelComp,razavikia2023SumCode,Yan2024Novel,yan2025remac,chen2025multi,razavikia2023computing,razavikia2024FedComp}, we proposed a new digital computation framework to design digital modulation for performing a general function computation. While this approach is broadly applicable, the algorithmic complexity required to optimize the modulation constellation grows rapidly with the constellation levels and with the number of nodes, thereby limiting its scalability for high-levels modulation or large-scale networks.

\subsection{Key Contributions of Part~II}

Part II introduces a pyramid sampling scheme to address these challenges: rather than designing the modulation based on all possible output points of the function to compute, we select a representative subset of function outputs. Under the assumption that the aggregation function is symmetric in each pre-processed value (output value of $\varphi_k(\cdot)$), we propose a pyramid sampling that reduces the computational cost of solving the optimization problem in \eqref{eq:main_problem} by orders of magnitude, at the expense of a controlled trade-off between complexity and accuracy for a given modulation level. In other words, increasing the sampling order raises computational load for designing the modulation but yields a more accurate modulation design, vice versa. Notably, it is shown that the special case of pyramid sampling with only $q$ samples avoids overlaps among the constellation points. At the extreme sampling order $p=K$, only those symbol combinations in which \emph{all} $K$ nodes agree are retained.  In this sampling scheme, termed as \emph{majority‐based sampling}, the receiver needs only to distinguish the $q$ points $f(z,z,\dots,z),~ z\in\{0,1,\dots,q-1\}$.
Since every node transmits the same constellation point $x_z$, the received sum  $r_z=\sum_{k=1}^K x_z \;=\; K\,x_z,$ guarantees that these $q$ sums never overlap in the constellation plane.  Consequently, the designer replaces an intractable $\mathcal{O}(q^K)$‐constraint optimization with a linear‐in‐$\mathcal{O}(q)$ problem, yet, by choosing a sufficiently high level of modulation, can in fact improve end‐to‐end computation accuracy compared to low modulation level exhaustive designs.

Moreover, many prior digital OAC schemes emerge naturally as special cases of majority‐based sampling. One-bit over-the-air aggregation method~\cite{zhu2020one}, for instance, corresponds to setting $q=2$, while frequency‐shift–keying vote‐aggregation arises by assigning each binary symbol to an orthogonal waveform~\cite{csahin2021distributed,csahin2023distributed}.  Extending the same principle to higher‐level phase-shift keying (PSK) or quadrature Amplitude Modulation (QAM) simply increases $q$, embedding all previously proposed digital OAC techniques within a single, unified sampling framework.  In this way, pyramid sampling offers both a principled complexity–accuracy trade‐off and a unifying perspective on the design of digital over–the–air computation methods.

Numerical results are provided to analyze the impact of sampling orders on the design of the constellation diagram and evaluate the computation performance of different digital modulations for OAC over a noisy MAC. This approach paves the way for scalable digital OAC in large‐scale edge networks, combining spectral efficiency with practical implementation. 

In summary, the main contributions of Part~II are as follows:
\begin{itemize}
    \item \textbf{Pyramid Sampling:} We introduce a novel pyramid sampling for digital OAC, which selects a subset of the superimposed constellation points to drastically reduce the encoder design complexity from $\mathcal{O}(q^K)$ to $\mathcal{O}(q^{K-p+1})$, where $p$ is the sampling order $p\in \{1,\ldots, K\}$. We also analyze the trade-off between optimization complexity and computation accuracy, i.e., how the sampling order raises the computational complexity for designing the modulation, and what the computation error is.

    \item \textbf{Majority-based sampling:} Based on the proposed pyramid sampling, we propose the majority‐based sampling method, which simply is the special case of pyramid when $p=K$ but with a high level of modulation $q'\gg q$. The notable feature of the majority‐based sampling is that it confines aggregation to $q$ constellation points, where they correspond to the majority of the node values, thereby inherently avoiding destructive overlaps. We show that it allows the use of all the existing digital modulation such as QAM, Hexagonal QAM,  PSK, ASK, BPSK, etc, to perform computing over the MAC. This proposed scheme is simple, but it can outperform all the state-of-the-art.   
    \item \textbf{Numerical experiments:}   Through extensive numerical simulations over various aggregation functions, modulation levels, and noise models, we demonstrate that moderate sampling orders achieve near‐optimal performance with orders‐of‐magnitude fewer constraints compared to exhaustive ChannelComp designs.
\end{itemize}

\subsection{Organization of the Paper}
The remainder of this paper is organized as follows. Section~\ref{sec:system_model} introduces the system model and recalls the digital OAC constellation design from Part~I. Section~\ref{sec:low-complexity} presents the pyramid‐based sampling framework and its complexity–accuracy trade‐offs. Section~\ref{sec:low-complexity} develops details for both pyramid and hard majority‐based sampling methods. Section~\ref{sec:Evaluation} provides numerical simulations, and Section~\ref{sec:Conclusion} concludes with directions for future work.

%---------------------------------------------------
\subsection{Notation}
%---------------------------------------------------

We denote by $\mathbb{N}$, $\mathbb{Z}$, $\mathbb{R}$, and $\mathbb{C}$ the sets of natural, integers, real numbers, and complex numbers, respectively. Scalars are represented by lowercase letters such as $x$, while operators are denoted using calligraphic letters like $\mathcal{A}$. For a complex number $x \in \mathbb{C}$, its real and imaginary parts are denoted by $\mathfrak{Re}(x)$ and $\mathfrak{Im}(x)$, respectively. The Euclidean norm of a vector $\bm{x} \in \mathbb{C}^n$ is denoted by $\|\bm{x}\|_2$, and its squared form is $\|\bm{x}\|_2^2 = \sum_i |x_i|^2$. For a set $\Omega$, we use $|\Omega|$ to present the cardinality of the set. For integers $n,k\in\mathbb{Z}$ with $0\le k\le n$, the binomial coefficient is denoted by $\binom{n}{k}=n!/(n-k)!k!$.   For an integer $N$, $[N]$ corresponds to the set  $\{1,2,\dots, N\}$.

The distribution $\mathcal{CN}(0,\sigma^2)$ denotes a circularly symmetric complex Gaussian distribution with independent real and imaginary components, each following a normal distribution $\mathcal{N}(0,\sigma^2/2)$.

\section{Communication Model and Problem Statement}\label{sec:system_model}

This part follows the same communication model as Part~I~\cite{razavi2025revisit}. Hence, we review them here; more details can be found in Part~I~\cite{razavi2025revisit}.  The communication model comprises the following key stages: 

1) \textit{Source Encoding and Quantization}: Each node \(k\) holds a scalar \(s_k\in\mathbb{R}\), which is mapped by a source encoder \(\varphi_k(\cdot)\) into a real codeword \(c_k\).  This is then quantized to \(q\) levels via a quantizer \(\mathcal{Q}\), yielding 
%------------------
\begin{align*}
    \tilde c_k \;=\; \mathcal{Q}\bigl(\varphi_k(s_k)\bigr) \in \mathcal{X}_k, 
\end{align*}
%------------------
where $\mathcal{X}_k$ denotes the set of alphabet for node $k$ with $q$ distinct values, where we assume to have the same cardinality as the level of the modulation, i.e., $|\mathcal{X}_k|=q$.

2) \textit{Digital Modulation and Superposition}: The symbol \(\tilde c_k\) is modulated by a common encoder \(\mathcal{E}:\mathcal{X}_k\to\mathbb{C}\) into a complex signal \(x_k\) with \(\lvert x_k\rvert^2\le1\) for $k\in [K]$.   All nodes transmit simultaneously over the MAC\footnote{We assume perfect synchronization among all the nodes and the CP. For imperfect synchronization, one may employ the existing techniques of OAC, e.g.,  blind schemes~\cite{saeed2022BlindFed,daei2025timely},  or waveform design~\cite{hellstrom2023optimal,evgenidis2024waveform} to manage the asynchronous transmission.}:
    \begin{equation}
      r= \sum_{k=1}^{K} p_kh_kx_k + z,
      \label{eq:channel_summary}
    \end{equation}
    where \(h_k\) is the channel gain, \(z\sim\mathcal{CN}(0,\sigma^2)\), and \(p_k = h_k^*/\lvert h_k\rvert^2\) implements channel‐inversion power control.  Hence, under ideal inversion
    %---------------
    \begin{equation}
      r = \sum_{k=1}^{K} \mathscr{E}_k(s_k) + z.
      \label{eq:channelfree_summary}
    \end{equation}
    %--------------
    In Part~I~\cite{razavi2025revisit}, we focus on designing the $\mathscr{E}(\cdot)$ given the distribution of the channel noise $z$. In this paper, we study the impact of sampling and quantization on the constellation design, which is given in the next step in the communication model.
    3) \textit{Decoding and Tabular Mapping}: The CP applies a maximum‐likelihood decoder \(\mathcal{D}(\cdot)\) to estimate
    \(\tilde{x} = \sum_{k} x_k\), then uses a tabular mapper \(\mathcal{T}:\mathbb{C}\to\mathcal{Y}_f\) to obtain
    \begin{align*}
        \tilde{f} = \mathcal{T}\bigl(\tilde{x}\bigr),
    \end{align*}
    which is finally sub-sampled via  \(\mathcal{Q}_p\) to produce the discrete output $ \hat{f} \in \mathcal{Y}_f^{p}$, where $\mathcal{Y}_f^{p}$ denotes the set of all possible output values for function $f$ after sampling with order of $p$. In practice, the tabular mapper $\mathcal{T}$ and $\mathcal{Q}_p$ act as one operator together, which directly maps the estimated values $\tilde{x}$ to the restricted range of the function $f$, $\mathcal{Y}_f^{p}$.  
     
     The overall communication model is depicted in Figure~\ref{fig:Systemmodel}. The main contribution of Part~II revolves around the idea of how to perform the sampling and quantization step after the received signal $r$, such that we can still compute $\hat{f}$ with adequate accuracy.  Moreover, this could help reduce the computational complexity cost for designing the encoder $\mathcal{E}$ in Step $2$.  The following section discusses the challenges and criteria in designing the decoder $\mathcal{T}(\cdot)$ and how they can be addressed.  
% -----------

% ==================
\input{Figs/Fig_ChannelComp_idea}
% ==================

\subsection{Problem Statement}\label{sec:problem_statement_summary}
The received signal in \eqref{eq:channelfree_summary} contains the superimposed constellation points of all $K$ nodes, $\sum_{k}\mathscr{E}(s_k)$, and it could result in overlaps among the output points of the encoder $\mathscr{E}(s_k)$ by different nodes and we obtain the same output values for multiple different output values of the function (see Figure~\ref{fig:range_vector}).  To avoid such overlaps, a margin-based formulation is introduced to enforce distinct function outputs mapping to distinct received sums~\cite{saeed2023ChannelComp}. To this end, we recall $\bm{x} := \bigl[x^{(1)}, \dots, x^{(q)}\bigr]^{\mathsf{T}}$ is the vector that contains all the constellation points, $x^{(q)}\in \mathbb{C}$,  produced by the encoder  $\mathscr{E}_k(\cdot)$. Also,  let \(\mathcal{Y}_g\) be the range of the function $g$, each the distinct outputs denoted by \(g^{(i)}\) for $i\in \{1,\ldots,|\mathcal{Y}_g|\}$.  For each index \(i\), define
\[
r_i \;=\; \sum_{\ell\in\mathfrak{I}_i} x^{(\ell)},
\]
where \(\mathfrak{I}_i\subset\{1,\dots,q\}\) is the combination of symbols yielding same \(g^{(i)}\).  We then solve
%-----------------
\begin{equation}
\label{eq:P2_summary}
 \max_{\bm{x},\,\lambda>0}\;\lambda~~
\text{s.t.}~~
\mathcal{D}_{\mathbb{C}}(r_i,\,r_j)\;\ge\;\lambda\,\bigl|g^{(i)}-g^{(j)}\bigr|^2,
\|\bm{x}\|_2^2\;\le\;1,
\end{equation}
%-----------------
for all $i\neq j, (i,j)\in \mathcal{Y}_g$,  where \(\mathcal{D}_{\mathbb{C}}\) is a distance metric in the complex plane. Also,  the unit‐power constraint \(\|\bm{x}\|_2^2\le 1 \) enforces normalization of the vector of constellation points.  This relaxation replaces the hard no‐overlap requirement in \cite[Eq~(8)]{razavi2025revisit} with a proportional‐margin constraint: maximizing \(\lambda\) embeds a Lipschitz‐smooth separation of constellation sums by the differences of \(g\) values. In Part~I, we studied the effect of the choice of distance $\mathcal{D}_{\mathbb{C}}$  on the constellation points. However, we still need to consider the computational complexity of solving the optimization problem in \eqref{eq:P2_summary}. Indeed, \eqref{eq:P2_summary}  imposes disproportionate computational complexity, which must be addressed.

The optimization in \eqref{eq:P2_summary} requires evaluating all function outputs, so the number of constraints grows exponentially with the number of nodes \(K\). In the worst case, the output space $\mathcal{Y}_f$ has \(q^K\) values, leading to \(\mathcal{O}(q^{2K})\) constraints. However, for practical functions such as sums or products, quantization limits the effective range to \(\mathcal{O}(q)\) values, since the result must fit a fixed bit-width in a digital system~\cite{choi2019accurate}.  To make the problem tractable, we restrict \(\mathcal{P}_2\) to a representative subset of quantized outputs. This compression maintains accuracy while keeping the optimization computationally feasible. However, for a given input quantization level \(q\) and network size \(K\), the key question becomes: which subset of outputs should be selected to minimize the computational complexity of solving \(\mathcal{P}_2\) while still achieving the required in the function computation when it has to be quantized by \(q\) quantization levels? In the next section, we introduce a simple yet effective pyramid sampling scheme that efficiently selects representative outputs, enabling tractable optimization while keeping the resulting error acceptably low.

\section{Pyramid Sampling for Symmetric Function }
\label{sec:low-complexity}

% *************************
\input{Figs/Fig_SquareConst}
% *************************

  Since we restrict ourselves to the class of the aggregation functions defined in Part~I~\cite{razavi2025revisit}, this aids us in designing the modulation only for the symmetry function $g(x_1,\ldots,x_K)$. Moreover, for ease of the presentation, we assume that the input values come from the same set of alphabet, i.e.,   $\mathcal{X}_1=\ldots=\mathcal{X}_K=\mathcal{X}$.   By invoking this symmetry, we can decompose the function $g$ by first computing the histogram from the input vector $(x_1,\ldots,x_K)\in \mathcal{X}^{K}$ and then applying $g$ as a post-processing map. Accordingly, we can define our sampling on the histogram of the input vector~\cite{zaheer2017deep}. In particular, designing the modulation for symmetry function $g$, allow us to approximately (due to quantization steps) compute all the functions $f: \mathbb{R}^{K}\mapsto \mathbb{R}$, that can be decomposed as $f =  g(\phi_1(s_1),\ldots,\phi_K(s_K))$, where $\phi_k(s_k): \mathbb{R} \mapsto \mathbb{R}$ and $g:\mathbb{R}^{K}\mapsto \mathbb{R}$ be a symmetry function. This class of function includes the class of nomographic functions, multiplicative nomographic functions, etc. 
  
  This pyramid sampling on the histogram considers a subset of all constraints in \eqref{eq:P2_summary} to be included for the design of modulation, which overall reduces the number of constraints to $\mathcal{O}(q^{p})$, where $p$ denotes the pyramid level.  More specifically, for a symmetric function $g$ with quantized input values $x_k\in \mathcal{X}$, we can define its input domain as a histogram. More precisely, let us consider a symmetric function, under the quantized input domain, i.e., 
  %--------
  \begin{align}
      g:\mathcal{X}^{K}\;\longrightarrow\;\mathcal{Y}_g,
  \end{align}
  %--------  
where \(|\mathcal{X}|= q\) and for some positive integer $q\in \mathbb{Z}^{+}$. without loss of generality, let $\mathcal{X}=\{0,1,\ldots,q-1\}$. By symmetry, \(g\) depends only on the histogram, i.e., let $\bm{h}:=(h_0,h_1,\dots,h_{q-1})$ be the histogram vector, where $h_j$ is the number of nodes that select input value $s_k$ to be $j$ for $j \in \{0,1,\ldots,q-1\}$, i.e., 
%--------
\begin{align*}
h_j=\bigl|\{k~|~s_k=j, k\in [K]\}\bigr|,  
\end{align*}
%--------
where $|\cdot|$ indicates the cardinality of the set, also, the validity of histograms mandates the equality between the number of nodes and the sum of histogram bins, i.e., $\sum_{j=0}^{q-1}h_j=K$. 
Hence, one may equivalently view $g:\Omega \mapsto \mathbb R,$ where  
%----------------
\begin{align}
\label{eq:Omega_define}
    \Omega=\{\bm h\in\mathbb{N}_{0}^q:\sum_j h_j=K\},
\end{align}
%----------------
 is set of all possible histogram vectors, where $\mathbb{N}_0$ denotes set of natural number with zero, i.e., $\mathbb{N}_0=\mathbb{N}\cup \{0\}$. Note that the  cardinality of $\Omega_p$ is $(\binom{K+q-1}{q-1}$, which grows as \(\mathcal O(q^{K})\). Then, based on this given histogram,  we define the pyramid sampling on this histogram in the following definition.
%*****************
\begin{definition}[Level-\(p\) Pyramid Sampling]
Let $g: \mathcal{X}^{K}\mapsto \mathbb{R}$ be a symmetry function, where $|\mathcal{X}|=q$ and $K$ is a positive integer indicating the number of input arguments. Also,  let $\Omega$ be the set of all possible histogram vectors, defined in \eqref{eq:Omega_define}. For a positive integer \(p\in\{1,2,\dots,K\}\), we define the sub-sampled histogram of level $p$ below.
\begin{align}
\nonumber \Omega_{p}
&=\,
\bigl\{\,
\bm h=(h_0^{p},\dots,h_{q-1}^{p})\in\mathbb{N}_0^q:\sum_{j=0}^{q-1}h_j^{p}=K,\\ \label{eq:sampling-set}
&h_j^{p}\in\{0,p,2p,\dots, K \}
\,\bigr\},
\end{align}
where $\Omega_1$ indicates the whole input domain, i.e., $\Omega_1 =\Omega$.  
\end{definition}
%*****************
It is not difficult to check that \(\lvert\Omega_p\rvert=\binom{K-p+q}{q-1} =\mathcal O(q^{K-p+1})\) for fixed \(q\) and \(p \in \{1,\ldots, K\}\). According to the definition of $\Omega_p$ in \eqref{eq;samplingOmegap}, the minimum variation among its elements is $p$. We depict an example of the pyramid sampling for $\Omega_2$ in Figure~\ref{fig:SqaureConst}.  To illustrate, consider the symmetric product function 
%--------------
\begin{align}
g(x_{1},x_{2}) = x_{1}x_{2},    
\end{align}
%--------------
defined over the alphabet \(\mathcal{X}=\{0,1,2,3\}\) with \(K=2\) and \(q=4\).  The histogram domain is
%---------------
\begin{align}
    \Omega = \bigl\{\,h=(h_{0},h_{1},h_{2},h_{3})\in\mathbb{N}_{0}^{4}:\sum_{j=0}^{3}h_{j}=2\bigr\},
\end{align}
%---------------
which has cardinality \(\binom{2+4-1}{4-1}=\binom{5}{3}=10\), and $g$ can be written as 
%---------------
\begin{align*}
    g(h_{0},h_{1},h_{2},h_{3})=0^{h_0}\times 1^{h_1}\times 2^{h_2}\times 3^{h_3},
\end{align*}
%---------------
where $h_{i}, \in \Omega$ for $i\in [3]$. Next, the level-\(p=2\) sampling on $\Omega$, gives us the following sub-sampled histogram. 
%------------
\begin{align}
    \Omega_{2}
\;=\;
\bigl\{\,(2,0,0,0),(0,2,0,0),(0,0,2,0),(0,0,0,2)  \bigr\},
\end{align}
%------------
whose size is
\(\binom{\tfrac{2}{2}+4-1}{4-1}=\binom{4}{3}=4\).  By symmetry the value of \(g\) depends only on the histogram \(h\) and can be written as
%-------------
\begin{align*}
    g(2,0,0,0)& =0\times 0, ~~g(0,2,0,0) =1\times 1,\\ ~~g(0,0,2,0) &=2\times 2, ~~ g(0,0,0,2) =3\times 3. 
\end{align*}
%-------------
 Consequently, designing or constraining \(\bar g\) on \(\Omega_{2}\) requires only four points instead of ten, which reduces the number of constraints by a factor of \(2.5\).

Then, by merging the sampling operator $\mathcal{Q}_p(\cdot)$ and tabular mapping operator $\mathcal{T}(\cdot)$ in Figure~\ref{fig:Systemmodel},  we introduce a pyramid sampling operator $\mathcal{T}_{p}:=\mathcal{Q}_p\mathcal{T}$ that samples the function output from  $\Omega_p$ by a given sampling order $p$, i.e., 
let \(\mathcal T_{p}\) be the projection (sampling) operator
\begin{align}
\label{eq;samplingOmegap}
    \mathcal{T}_{p}:=
    \;g\big\vert_{\Omega}
\;\longmapsto\;
\hat g
\;=\;
g\big\vert_{\Omega_p},
\end{align}
so that \(\hat g:\Omega_p\to\mathcal{Y}_g^{p} \), where $\mathcal{Y}_g^{p}\subseteq \mathcal{Y}_g$ is the range of function $g$ retaining  only those histogram‐elements in \(\Omega_p\). Since the function  $g: \Omega \to \mathbb{R}$ is symmetric, it is not difficult to see that for the case when $p=1$, we have  \(\Omega_1=\Omega\), the operator becomes an identity operator, i.e., 
%--------------
\begin{align}
   g = \mathcal{T}_{{1}}(g) = \mathcal{I}(g),\quad \Omega_{1} = \mathcal{X}^{K}.
\end{align}
%--------------
Also, for \(p=K\), we get \(\Omega_K=\{\,K\bm e_j : j=0,\dots,q-1\}\) has size \(q\), yielding the \emph{coarsest} quantization. Indeed, we consider $\Omega_K$ for the sampling which gives us $q$ levels, i.e., 
%--------------
\begin{align}
    \hat{g} = \mathcal{T}_{K}(g),
\end{align}
%--------------
While such sub-sampling from $q^K$ to $q$ (\(\lvert\Omega_K\rvert=q\ll \binom{K+q-1}{q-1}\)) is quite coarse and loses a lot of information, the key advantage is that now we can set $q$ to be so high that the error is negligible. Then, to quantify the computational complexity of designing the modulation, we consider the log of the cardinality of $\mathcal{Y}_g^{p}$, i.e.,  
%----------------
\begin{align}
\label{eq:delta_definition}
\delta(p):={\log(\lvert\mathcal{Y}_g^{p}\rvert)}, \quad p\in [K]. 
\end{align}
%----------------
Note that we always have $|\mathcal{Y}_{g}^{p}|\leq |\Omega_p|$ for a general symmetry function $g$, whose the upper bound $\log(|\Omega_p|)$ for fixed \(q\) and \(p\ll K\) scales as \(\mathcal{O}({(K-p+1)}\times \log{q})\). Also, $\delta$ scales inversely with sampling order-$p$ and we have
%------------
\begin{align}
    0\leq  \delta(p) \leq \log(|\Omega_1|), \quad p\in[K], 
\end{align}
%------------
where zero can be attained for a constant function, i.e., $g=c, c\in \mathbb{R}$.  For a general symmetry function, $\delta(p)= \log(|\Omega_p|)$, for $p\in [K]$.  To capture the computation error induced by this sampling scheme, we define the normalized sampling error below. 
%--------------
 \begin{align}
 \label{eq:epsilon_definition}
     \epsilon(p)
\;:=\;\frac{1}{K}
\max_{\bm h\in\Omega}
\min_{\bm h'\in\Omega_p}
\bigl\lvert g(\bm h)-g(\bm h')\bigr\rvert, \quad p \in [K]. 
 \end{align}
%--------------
This metric implies worst-case sampling error.  For instance, suppose \(g\) is \(L\)-Lipschitz under the \(\ell_\infty\)–norm on histograms. Then, it directly gives us 
%--------------
\begin{align}
     \frac{L}{K}=\epsilon(1)\leq \epsilon(p)\leq \epsilon(K)=L, \quad p\in [K]. 
\end{align}
%--------------
We also listed the values of $\epsilon(p)$ for different symmetry functions in Table~\ref{tab:epsilon}.  Therefore, given the sub-sampled histogram $\Omega_p$, we have a trade‐off between complexity for the pair of $(\epsilon, \delta)$ depending on the sampling order $p$, since increasing $p$ results in lower complexity $\delta$, and higher computational error~$\epsilon$. 

%*****************
\input{Figs/Fig_table_epsilon}
%*****************

In Figure~\ref{fig:tradeoff}, we plot the normalized computational costs, i.e., \(\delta(p)/\log{|\Omega_1|}\) against  \(\epsilon(p)\) for
\(q\in\{4,32,64,128\}\) with \(K=100\). As \(p/K\) increases even
modestly, \(\delta(p)\) drops sharply. Further error allowance yields little extra reduction in complexity. Larger \(q\) begins at higher complexity but also decays faster, offering
flexibility in balancing cost versus accuracy.

%*****************
\input{Figs/Fig_tred_off}
%*****************

We note that if the input values $\bm{s}_k$ comes form a compact space, e.g.,   $\bm{s}_k\in [0,1]$ for $k\in [K]$. Then, $\epsilon(p)$ for \(L\)-Lipchitz function $g$ can be upper bound by $Lp/q$. This implies that we can reduce the normalized sampling by increasing $q$.  Accordingly, a higher order of sampling $p$ can be employed, and then the error can be computed and reduced with a higher level of input quantization $q$.
To enable a fair comparison, consider there computation settings: (i) a baseline with full quantization at resolution $q^K$, denoted $\Omega_1$, with no sampling; and (ii) a sampled approach with sampling order $K$, where the effective quantization is $q' = q^K$, denoted $\Omega_K$. (iii) an intermediate configuration with sampling order $K$ and quantization at resolution $q \times K$.   Settings (i) and (ii) incur comparable computational costs in the design of the modulation vector, whereas settings (i) and (iii) offer similar levels of computational accuracy. The following proposition formalizes this comparison.

\begin{prop}\label{prop:quantization}
Let \(g:\Omega\to\mathbb R\) be symmetric and \(L\)-Lipschitz under the \(\ell_\infty\)–norm on histograms of \(K\) inputs drawn from \([0,1]\), each quantized into \(q\) levels.  Let \(\delta(p)\) and \(\epsilon(p)\) be as in \eqref{eq:delta_definition} and \eqref{eq:epsilon_definition}. Then, consider the following three scenarios:
\begin{itemize}
    \item Let the level of the modulation be $q$, and sampling order $p=1$, where we use $\epsilon_{\rm{ChannelComp}}$ and $\delta_{\rm{ChannelComp}}$ be the normalized sampling error and computational cost. 
    \item  Let $\epsilon_{\rm{Pyramid,1}}$ and $\delta_{\rm{Pyramid,1}}$ be the normalized sampling error and computational cost for the case, where  $q^K$ is the level of the modulation and sampling order $p=K$. 
    \item Let $\epsilon_{\rm{Pyramid,2}}$ and $\delta_{\rm{Pyramid,2}}$ be the normalized sampling error and computational cost for the case, where  $q\times K$ is the level of the modulation and sampling order $p=K$. 
\end{itemize}
 Then, we have
%----------
\begin{subequations}
\begin{align}
    & \frac{\epsilon_{\rm{Pyramid,1}}}{\epsilon_{\rm{ChannelComp}}}=\frac{K}{q^{K-1}},
~
\frac{\delta_{\rm{Pyramid,1}}}{\delta_{\rm{ChannelComp}}}\sim \mathcal{O}\Big( \frac{K}{(q-1)\log_{q}\!({K})}\Big), \\
 &\frac{\epsilon_{\rm{Pyramid,2}}}{\epsilon_{\rm{ChannelComp}}}=1,
~
\frac{\delta_{\rm{Pyramid,2}}}{\delta_{\rm{ChannelComp}}}\sim \mathcal{O}\Big( \frac{1}{q-1}\Big),
\end{align}
\end{subequations}
%---------- 
for $K\gg q$. 
\end{prop}
% %***********************

\begin{proof}
See Appendix~\ref{Ap:Proof_comelxity}.
\end{proof}

% ***********************
\input{Figs/Fig_Qunatization_plot}
%************************

The first scenario in Proposition~\ref{prop:quantization} is implemented by ChannelComp in \cite{saeed2023ChannelComp}, whereas the second and third scenarios reflect the sampling error inherent in our proposed scheme. Indeed, the second scenario with $(\epsilon_{\rm{Pyramid,1}},\delta_{\rm{Pyramid,1}})$, we obtain higher computational complexity for cases where $K\gg q$ with significantly lower computational error compared to ChannelComp. The third scenario with $(\epsilon_{\rm{Pyramid,2}},\delta_{\rm{Pyramid,2}})$ considers the case where we obtain similar computational error while resting in much lower computational complexity compared to ChannelComp, for large level of modulation, $q\gg1$. Consequently, Proposition~\ref{prop:quantization} demonstrates that applying high-level modulation with a high level of sampling to both input and output values can either significantly reduce computation error or computational cost for solving the optimization compared to employing low-precision quantization solely for the input.

%--------------
\input{Figs/Fig_TriangleCont}
%--------------

In Figure~\ref{fig:quantize}, we consider only the computation aspect of Proposition~\ref{prop:quantization}. In particular, Figure~\ref{fig:quantize} presents the Monte Carlo–estimated MSE for the sum and product functions with $K=5$ inputs, comparing three sampling schemes: 1) sampling of order $p=1$ with quantization levels $q$, $p=K$ with quantization level $qK$, and sampling of order $p=K$ with quantization level of $q^K$.  Over $2\times 10^3$ random samples in $[0,1]^5$, the high‐levels scheme ($q^K$) drives the MSE down by up to two orders of magnitude relative to simple input quantization ($p=1$), while the intermediate scheme ($qK$) delivers a proportional error reduction.  We note that these results assume an ideal communication channel and do not account for channel noise.

\subsection{Resolving Overlaps for \texorpdfstring{$\Omega_p$}{Omega}}

To avoid overlaps among the constellation points induced by the set $\Omega_p$, we must ensure that each pair of constellation points in the set satisfies the distance constraint in \eqref{eq:P2_summary}. Regarding the overlaps, we have the following proposition.

% --------------
\begin{prop}
 \label{prop:ComQuntize}
For $p \in \{1,\ldots,K\}$, let $\Omega_p$ be the sampling set for the symmetry function $g$ with sampling operator $\mathcal{T}_{p}$ as defined in \eqref{eq;samplingOmegap}.  To avoid any overlaps among the constellation points of $\hat{g} = \mathcal{T}_{p}(g)$, it is sufficient that the superimposed constellation points satisfy the following condition   
% --------------
\begin{align}
  \label{eq:d_x_relax2}
    \mathcal{D}_{\mathbb{C}}(r_i,\,r_j) \geq  \max_{(\bm{x}^{(i)},\bm{x}^{(j)})\in \Omega_{p'}}  |\hat{g}(\bm{x}^{(i)}) - \hat{g}(\bm{x}^{(j)})|^2,
\end{align}
% --------------
for any $p'\leq p$.  Here, $(r_i,r_j)$ are the constellation points correspond to the input value $\bm{x}^{(i)}$ and $\bm{x}^{(j)}$, respectively. 
\end{prop}
% --------------
\begin{proof}
    For the $p'=p$, this becomes identical to \cite[Lemma~1]{saeed2023ChannelComp}, therefore, \eqref{eq:d_x_relax2} avoids overlaps among the points in $\Omega_p$. For $p'\leq p$, we have $\Omega_p \subset \Omega_{p'}$ by definition in \eqref{eq:sampling-set}. Accordingly, satisfying  \eqref{eq:d_x_relax2} for $p'<p$, also prevents destructive overlaps among the points $ \Omega_{p}$. 
\end{proof}

Based on Proposition~\ref{prop:ComQuntize}, if we solve the optimization problem in \eqref{eq:main_problem} for $\Omega_p$, the resultant constellation points would also support a lower resolution $\Omega_{p'}$ with $p'\geq p$, although not the reverse. Essentially, we have to increase the cardinality subset (lowering the value of $p$) and ensure satisfaction of all the constraints in \eqref{eq:d_x_relax2}. 

However, in the case of limited computational resources,  we can slightly mitigate this issue by capturing some redundancy in this sampling set.  For instance, the changes that follow a linear trajectory are proportional to each other with a factor that depends on the choice of distance $\mathcal{D}_{\mathbb{C}}$ (e.g., see Figure~\ref{fig:Trinagle}). Hence, we can only consider the maximum of them. Specifically, for every pair $(i,j)$ of the transmitted constellation points, we define the subset $ \mathcal{S}^{(p)}_{i,j}$ as the collection of output pairs induced by a transition in $p$ node’s input from state $i$ to state $j$. In contrast, the remaining nodes maintain their respective values. Formally, for a two quantized input values $\bm{c},\bm{c}'\in \mathcal{X}^{K}$ with $|\mathcal{X}|=\{1,\ldots,q\}$, let
% --------------
\begin{align}
   \nonumber
    \mathcal{S}^{(p)}_{i,j} = \{ (\tilde{\bm{c}},\tilde{\bm{c}}') & : |\epsilon(\tilde{\bm{c}},\tilde{\bm{c}}')| = p \text{and}
     \forall k \in \epsilon(\tilde{\bm{c}},\tilde{\bm{c}}'), \\ &(\tilde{c}_k,\tilde{c}'_k) = (i,j) \}. \label{eq:Sij_define}
\end{align}
% --------------
for $(i,j)\in [q]\times[q]$ and $p\in [1,\cdots,K]$. These sets capture a subset of sets ${\Omega}_{p}$ (see Figure~\ref{fig:Trinagle}), i.e.,  for $p \geq 1$, we have
%---------------
\begin{align}
       \bigcup_{p=1}^{K} \bigcup_{(i,j)}\mathcal{S}^{(p)}_{i,j} \subseteq {\Omega}. 
\end{align}
%---------------
Then, the constraints in \eqref{eq:d_x_relax2} for each pair $(i,j)$, we can be rewritten as
% --------------
\begin{align}
  \label{eq:d_x_relax}
     \mathcal{D}_{\mathbb{C}}(r_i,r_j)\geq\max_{\ell \in [p,2p,\ldots, K]} \kappa_{\ell}\max_{(\tilde{\bm{x}},\tilde{\bm{x}}')\in \mathcal{S}^{(\ell)}_{i,j}}  |g(\tilde{\bm{x}}) - g({\tilde{\bm{x}}'})|^2, 
\end{align}
% --------------
where $\kappa_{p} := \mathcal{D}_{\mathbb{C}}(x_1,x_2)/ \mathcal{D}_{\mathbb{C}}(p x_1,p x_2)$, for all $p \in \{1,\ldots,K\}$. In this way, we pick only the maximum overlaps among the constellation points along a linear trajectory.  

Regarding the cardinality of the sum set of $\mathcal{S}^{(p)}_{i,j}$, we have the following lemma.
%--------------
\begin{lem}\label{lem:cardinal}
    Let $\mathcal{S}^{(p)}=\bigcup_{p=1}^K\mathcal{S}^{(p)}_{i,j}$ for $(i,j)\in [q]\times[q]$, where $\mathcal{S}^{(p)}_{i,j}$ as defined in \eqref{eq:Sij_define}.  Then, the cardinality of $\mathcal{S}_{i,j}$ is given by 
    %---------------
    \begin{align}
        |\mathcal{S}_{i,j}| = \binom{K+q-1}{q}, \quad (i,j)\in [q]\times[q].
    \end{align}
    %---------------
\end{lem}
\begin{proof}
    For the proof, see Appendix~\ref{ap:cardinal}. 
\end{proof}
%--------------
 Consequently, we reduce the number of all constraints corresponding to a linear trajectory in the optimization problem in \eqref{eq:main_problem} by a factor of $1/|\mathcal{S}_{i,j}|$ (only the maximum one is included in the optimization problem). 

We observe that the constraints in \eqref{eq:d_x_relax} alone do not guarantee the absence of overlap.  To illustrate this, consider the constellation diagram in Figure~\ref{fig:Trinagle}, where the MAC yields six points \(\{r_{1},r_{2},\dots,r_{6}\}\).  Constraint \eqref{eq:d_x_relax} enforces minimum distances only for the pairs linked by solid blue lines, while the red dashed‐line pairs 
\(\{r_{1},r_{5}\}\), \(\{r_{3},r_{4}\}\), and \(\{r_{2},r_{6}\}\) remain unconstrained and may therefore coincide.  In practice, such coincidences are rare, since the solid‐line constraints typically suffice; preventing them would require a non-regular mapping \(g\).  Indeed, for such a case, the only mapping that produces an overlap is the pathological example in \eqref{eq:counterexmaple}, which forces the following values.   
%----------------
\begin{subequations}
    \label{eq:counterexmaple}
    \begin{align}
    g(0,0) &= \mathcal{T}(r_1) = b, \\
    g(0,1) &= g(1,0) = \mathcal{T}(r_4) = a, \\
    g(0,2) &= g(0,2) = \mathcal{T}(r_2) = a, \\
    g(1,2) &= g(2,1) = \mathcal{T}(r_5) = 2a, \\
    g(1,1) &= \mathcal{T}(r_6) = b, \\
    g(2,2) &= \mathcal{T}(r_3) = b,
    \end{align}
\end{subequations}
%----------------
for two nonzero real value numbers $a\neq b,a,b\in \mathbb{R}$, thereby assigning two different values to the same point. However,  such a function is not usually encountered in standard over‐the‐air computation literature and lacks practical significance.  Moreover, after sampling reduces the constellation to \(\{r_{1},r_{3},r_{6}\}\), no overlaps remain.

\begin{rem}\label{rem:suminvarince}
   We note that for the sum function and the choice of the Euclidean distance for the computation condition in Proposition~\ref{prop:ComQuntize}, i.e.,  
   %-----------
   \begin{align}
   \mathcal{D}_{\mathbb{C}}(r_i,r_j) = |r_i-r_j|^{2},
   \end{align}
   %-----------
   which is the proposed distance by the ChannelComp in \cite{saeed2023ChannelComp}. Following similar steps as in \cite{razavi2025revisit}, it is straightforward to show that the solution to the optimization problem in \eqref{eq:main_problem} is invariant to the value of $p$. Indeed, regardless of the value of $p$, we always get the PAM modulation as the optimal solution. 
\end{rem}

\subsection{Majority-Based Sampling without Overlaps}

In this subsection, we study a special case of the proposed sampling scheme, where the sampling order is $p=K$, resulting in a majority-based sampling scheme with set $\Omega_K$. Indeed,  a salient feature of this high-order sampling strategy is to prevent overlap among the constellation points over the MAC, provided each node selects its symbols from a set of \(q\) distinct constellation points. Consequently, the problem of signal overlap at the receiver is inherently avoided. Moreover, the shape constellation points induced by $\Omega_K$ are the same as the individual node constellation diagram, except that everything is scaled by a factor $K$. This feature helps us design the modulation with much less computation and in a more straightforward way. We highlight that any digital modulation format—QAM, hexagonal QAM, PSK, etc.—can be used without overlap concerns. Raising the modulation level \(q\) further improves computation accuracy by reducing quantization error. Moreover, by tailoring the constellation design to the specific aggregation function, we can fully exploit the advantages of digital OAC, achieving maximum performance gain under channel noise and implementation constraints.

Indeed, under this majority-based scheme, the OAC problem from the receiver's perspective can be interpreted as a point-to-point communication problem. All transmitters act collectively to vote on one of the \(q\) constellation points. The key distinction is that the received signal encodes the output of a desired function rather than a transmitted symbol.

\begin{prop}
Let  $ \bm{x}= [x^{(0)},x^{(1)},\dots,x^{(q-1)}]^{\mathsf{T}}\in \mathbb{C}^{q},$ be a \(q\)-ary constellation with
$x^{(i)} \neq x^{(j)}$ for all $i\neq j$. Each node \(k=1,\dots,K\) quantizes its input \(s_k\in\mathbb{R}\) by
$\tilde c_k = Q(s_k)\in\mathcal{X}$ with $|\mathcal{X}|=q$. For input vector $\bm{c}\in \mathbb{R}^{q}$,  the define the majority operator $\mathcal{M}_q:\mathbb{R}^{K}\mapsto \mathbb{R}^{K}$ as follows. 
%--------------
\begin{align}
    \mathcal{M}_g(\tilde{\mathbf c})
=\begin{cases}
g(\bm{a}_i) ~~~~~\bigl|\{k:\tilde c_k=a_i\}\bigr|
>\max_{j\neq i}\bigl|\{k:\tilde c_k=a_j\}\bigr|,\\[5pt]
\frac{1}{r}\sum_{j}g(\bm{a}_{i_j}), ~\text{if a tie occurs among }r\text{ points},
\end{cases}\label{eq:majority_samples}
\end{align}
%--------------
where $\bm{a}_i=(a_i,\ldots, a_i)\in \mathbb{R}^{K}$ with $a_i\in \mathcal{X}$ for $i \in [q]$,  and the symmetry function $g$ can be sampled by the corresponding sampling operator, i.e., 
\begin{align}
\label{eq:majority_operator}
    \mathcal{T}_{K}(g)(\mathbf{s})
:=
\mathcal{M}_g\bigl(\mathcal{Q}(\bm{s})\bigr),\quad  
\end{align}
  Under this setting, no overlaps exist among the superimposed constellation points.
\end{prop}
\begin{proof}
By construction, whenever the majority index is \(i\), and there is no tie among the points,  all \(K\) terms in the sum \(\sum_{k=1}^K\tilde c_k\) equal \(x^{(i)}\), so
\begin{align}
    r_i \;=\;\sum_{k=1}^K x^{(i)}
       \;=\; K\,x^{(i)}, \quad i\in [q]. 
\end{align}
Since the original constellation points \(x^{(i)}\) are distinct, scaling by the nonzero factor \(K\) preserves distinctness; for \(i\neq j\) we have 
\begin{align}
      r_i - r_j 
  = K\bigl(x^{(i)}-x^{(j)}\bigr)
  \;\neq\;0.
\end{align}
Hence the \(q\) possible superimposed symbols \(r_0,\dots,r_{q-1}\) are all different.  The tie points are exactly on the boundaries between two extreme points, and we have assigned the average values as the output values.    
\end{proof}

In Figure~\ref{fig:majority}, we depict an example for this sampling scheme, for $q=4$ and the desired function is the product function $f(s_1,x_2)=s_1s_2$, where $s_1,s_2\in \{0,1,2,3\}$. Since we only consider the corner points of the superimposed constellation diagram, the overlaps at the center points between the output $2$ and $0$ would not affect the computation procedure at the cost of losing accuracy.

%****************
\input{Figs/Fig_majority}
%****************

%****************
\input{Figs/Fig_table_compare}
%****************

%***********
\input{Figs/Fig_Num1}
%***********

\begin{rem}
    We note that the one-bit over-the-air computation scheme in~\cite{zhu2020one} corresponds to the special case of this majority sampling scheme when \(q=2\) and the desired function $g$ is the  Arithmetic mean function with $\mathcal{X}=\{-1,1\}$. Moreover, if the quantization code-book is considered a balanced number system, then for the averaging function, this scheme becomes similar to the majority vote computation scheme proposed in \cite{csahin2023over} with the desired function to be the  Arithmetic mean function. However, instead of the frequency, the voting happens over the set of constellation points. 
\end{rem}
In Table~\ref{tab:compraison}, we provide an overview of the sampling regimes—full enumeration at \(p=1\), intermediate pyramid sampling for \(1<p<K\), and majority-based sampling at \(p=K\)—by listing the desired function \(g\), the corresponding operator \(\mathcal{T}_{p}\), and the method used.

For the case where the function $g$ is the Arithmetic mean function, i.e., $g(s_1,\ldots,s_K)=\sum_{k}^Ks_k/K$, it is interesting that this case can be translated as a standard point to point communication scheme, where the role of nodes is to vote for the consensus constellation point to transmit over the MAC. Precisely,  the output of the $\mathcal{T}_{K}(g)(\mathbf{s})$  becomes the input values $s_k\in \{0,1,\ldots,q-1\}$ and the operator $\mathcal{T}_{K}$ acts as a majority voter on the input values, i.e.,  
%------------
\begin{align}
    \mathcal{M}_g(\tilde{\mathbf s})
=\begin{cases}
i ~~~~~\bigl|\{k:\tilde c_k=i\}\bigr|
>\max_{j\neq i}\bigl|\{k:\tilde c_k=j\}\bigr|,\\[5pt]
\frac{1}{r}\sum_{j}i_j, ~\text{if a tie occurs among }r\text{ points},
\end{cases}\label{eq:majority_samples_2inpute}
\end{align}
%------------
Therefore, CP decodes a value from $r\in \{0,1,\ldots,q-1\}$, the same as each node's alphabet.  All transmitters utilize a common codebook $\mathcal{X}$, which is also known to the CP. Importantly, the role of the transmitters is limited to collectively determining which symbol from the codebook is to be transmitted; in fact, a single transmitter can transmit the consensus constellation points on behalf of all others without affecting the correctness of the computation at the receiver. From the CP's perspective, the communication appears indistinguishable from standard point-to-point transmission.  

Since the communication over the MAC is reduced to standard point-to-point communication, one can use all the existing schemes for designing the constellation diagram in the literature and obtain the optimum constellation diagram based on some criteria~\cite{huber1994codes,laroia1994optimal}.  For instance, following Shannoon Kotelinkove's map, we can use Spiral-like curves to transmit the information, which is optimal for analog coding over an AWGN channel~\cite{akyol2010optimal}.

\section{Performance Evaluation}\label{sec:Evaluation}

This section presents a comprehensive set of numerical experiments to validate and compare the proposed constellation design and sampling strategies for different sampling leaves and modulation.  First, we investigate how pyramid sampling orders \(p\) affect the output of the SDP optimization design and the resulting constellations for different aggregation functions.  Next, we assess the end‐to‐end accuracy of the majority‐sampling scheme across multiple quantization sizes \(q\in\{16,32,64\}\) and modulation formats under AWGN.  Finally, we directly compare the SDP‐designed ChannelComp constellations against the majority‐sampling approach for the sum, product, and max functions, highlighting the trade‐offs between design complexity and computation accuracy. Throughout this section, we measure performance via the mean-squared error, i.e., 
%--------------
\begin{align*}
    \mathrm{MSE}(f)
=\frac{1}{N_s}\sum_{j=1}^{N_s}\bigl|f(\bm{s}_j)-\hat f_j\bigr|^2,
\end{align*}
%--------------
where \(f(\bm{s}_j)\) is the true function on the continuous input and \(\hat f_j\) its reconstruction, and $N_s$ is the number of Monte Carlo trials.

%***********
\input{Figs/Fig_Num2}
%***********

\subsection{Pyramid sampling orders}

This experiment examines the impact of the sampling order \(p\) and the corresponding sampling set \(\Omega_p\) on the solution of the max–min semidefinite program (SDP) in \eqref{eq:main_problem}, as well as on the resulting constellation diagrams for the desired functions $f_1 = \max_{k=1}^K s_k$ and  $f_2 = \prod_{k=1}^K s_k$.  We fix the quantization level size to \(q=8\) and the dimensionality to \(K=16\), and adopt the Euclidean distance metric for \(\mathcal{D}_{\mathbb{C}}\) in the SDP. The optimization is performed under four pyramid‐shaped sampling regimes, namely $p \in \{1,2,4,8\}$. The resultant constellation plots are depicted in Figure~\ref{fig:constellation-pyramid}.

As predicted by Remark~\ref{rem:suminvarince}, the solution for the sum‐based criterion always yields a PAM constellation, independent of \(p\). In contrast, the product‐based criterion (bottom row of Figure~\ref{fig:constellation-pyramid}) exhibits no point collisions when \(p=1, 2\), corresponding to the full index set \(\Omega_1\). With increasing \(p\), the first two constellation points (indexed by \(s_k=1,2\)) progressively converge until they coincide entirely at \(p=8\). Notably, $\lvert \Omega_8\rvert = \mathcal{O}(q^2)$ and $\lvert \Omega_1\rvert = \mathcal{O}(q^{16})$. Thus, increasing the sampling order yields a dramatic reduction in computational complexity—dropping from \(\mathcal{O}(q^{16})\) to \(\mathcal{O}(q^2)\)—at the expense of a single overlap, while preserving the overall geometric structure of the constellation.

%***********
\input{Figs/Fig_Num3}
%***********

\subsection{Majority-based Sampling with Different Modulation}

To evaluate the end-to-end accuracy of the extreme-point decoding under both quantization and AWGN, we draw \(N_s=10^5\) independent samples \(s\sim\mathcal U(0,1]\), shared across \(K=50\) nodes.  Each sample is uniformly quantized into \(q\in\{16,32,64\}\) levels with midpoints $m_i={(i+0.5)}/{q},\quad i=\{0,\dots,q-1\}$. 
Then it is mapped to one of three unit-power constellations (rectangular QAM,  PAM, or hexagonal) of level \(q\).  The \(K\) transmitted symbols are summed, corrupted by complex AWGN at per-node SNR from $0$ dB to $20$ dB, and demodulated by selecting the nearest scaled point \(K\cdot x\).  From the recovered midpoint \(\hat m\) we form the estimates  
\(\hat f_{1}=(\prod_k s_k)^{1/K}\) 
and \(\hat f_{2}=\max_k s_k\).  Figure~\ref{fig:mse-comparison} plots MSE versus SNR for each desired function.  At high SNR, all curves saturate at the uniform-quantizer distortion, which decreases as \(q\) increases.  At low SNR, AWGN-induced decoding errors dominate; rectangular QAM yields the lowest error due to its larger minimum distance, followed by hexagonal and then PAM.  The Geometric mean function exhibits the steepest MSE growth, since errors in \(\hat f\) are amplified by the \(K\)th power, whereas the max function grows more moderately (with max tracking the quantization MSE directly and sum scaling by \(K\)).  These results confirm that sufficiently fine quantization and constellation geometry jointly govern the overall computation accuracy.

\subsection{ChannelComp versus Majority Based Sampling}

This experiment evaluates the SDP‐designed constellations from Section~\ref{sec:low-complexity} under strict computational and channel constraints. For each desired function \(f\in\{\sum_k,\prod_k,\max_k\}\), continuous inputs \(s_k\in(0,1]\) are quantized into those quantized levels, mapped to the optimized symbols from \eqref{eq:main_problem}. To obtain the constellation diagram, we consider two scenarios, where we solve the max–min SDP to place \(q_1=4\) and $K=10$, and the other one where \(q_2=1024\) and $K=2$, which results in similar computational cost. However, we perform the transmission for $k=10$ nodes.  Each node transmits over a complex AWGN channel with per‐node SNR swept from $0$ dB to $20$ dB.  Maximum likelihood decoding then recovers each transmitted level.  Figure~\ref{fig:mse2-comparison} depicts the end‐to‐end distortion for all three functions, showing that in high SNR regions, the high resolution level scheme $q_2$ a outperform the low resolution scheme: the sum and max functions for low resolution scheme rapidly reach their quantization floors at moderate SNR, whereas the product mapping incurs larger errors at low SNR. Overall, the second scenario, where we perform the quantization at a much higher level, results in significantly lower computational error, which is spotted by Proposition~\ref{prop:quantization}.

\section{Conclusion}\label{sec:Conclusion}

This paper has introduced a pyramid sampling framework for digital OAC that reduces encoder‐design complexity from $\mathcal{O}(q^K)$ to $\mathcal{O}(q^{K-p+1})$ by selecting a representative subset of superimposed constellation sums, where $p$ is the sampling order and $K$ denotes the number of nodes. Under symmetric aggregation, this strategy offers a tunable trade‐off between computational burden and reconstruction accuracy. In the extreme case of majority‐based sampling ($p=K$), aggregation is confined to $q$ consensus points, inherently avoiding destructive overlaps and permitting the use of standard modulations (e.g., QAM, PSK). We proved that this majority‐based scheme inherently avoids any destructive overlaps, since each received sum is a scaled version of a single constellation point.   This unifies and generalizes prior OAC methods—one‐bit aggregation, balanced‐vote schemes—within a single framework. Furthermore, extensive simulations across sum, product, and maximum functions, various modulation levels, and noise models demonstrate that moderate sampling orders achieve a fair accuracy with orders‐of‐magnitude fewer constraints. Overall, the proposed approach enables scalable, spectrally efficient, and practically implementable digital OAC in large‐scale edge networks.

Future work will focus on constellation design for majority-based sampling, accounting for quantization effects and channel noise. A compelling question is whether optimal constellations can be derived for specific function classes under majority-based sampling. Given the alignment between majority-based sampling and conventional communication strategies, deriving such constellations may be tractable. Another direction involves applying the proposed sampling strategy to machine learning applications—such as distributed and federated learning—and evaluating its impact on model convergence and learning performance.

\appendix
\subsection{Proof of Lemma~\ref{lem:cardinal}}\label{ap:cardinal}

The superimposed constellation diagram can be seen as a Mikowski sum of each constellation diagram; this problem can be bijectively mapped to the combinatorial balls and bins problem~\cite{mitzenmacher2017probability}, where we seek the number of possible ways to put $K$ indistinguishable balls into $q$ distinguishable bins. For the set $\mathcal{S}_{(i,j)}^{(p)}$, we have already occupied $p$ positions, and the other positions need to be selected. Therefore,   this is precisely the number of integer solutions to 
% --------------
\begin{align}
    n_0+n_1+\cdots+n_{q-1}=K, n_i\geq 0.
\end{align}
% --------------
This gives us the following combinatorial counting.  
% --------------
\begin{align}
    |\mathcal{S}_{(i,j)}^{(p)}| = \binom{K+q-1-p}{q-1}, \quad p \in [K]. 
\end{align}
% --------------
Since by definition we know that $\mathcal{S}_{(i,j)}^{(p)} \cap \mathcal{S}_{(i,j)}^{(p')}=\emptyset$ for $p \neq p'$, we have 
% --------------
\begin{align}
     \nonumber
    |\mathcal{S}^{(i,j)}| & = \sum_{p=1}^K|\mathcal{S}_{(i,j)}^{(p)}| = \sum_{p=1}^K\binom{K+q-1-p}{q-1}, \\\nonumber
    & = \sum_{m=0}^{K-1}\binom{m+q-1}{q-1}= \binom{K+q-1}{ q},
\end{align}
% --------------
where the last equality comes from hockey-stick identity\cite{jones1996generalized}. Hence, we can conclude the proof. 

\subsection{Proof of Proposition~\ref{prop:quantization}}\label{Ap:Proof_comelxity}
    The ratios for the sampling errors come directly by substituting their corresponding value in the definition of $\epsilon(p)$ in \eqref{eq:epsilon_definition}. For the normalized computational costs, the values are given by
    \begin{align}
        \frac{\delta_{\rm{Majority,1}}}{\delta_{\rm{ChannelComp}}} &= \frac{K\log{q}}{\log{\binom{K-1+q}{q-1}}}, \\
         \frac{\delta_{\rm{Majority,2}}}{\delta_{\rm{ChannelComp}}} & = \frac{\log{(Kq)}}{\log{\binom{K-1+q}{q-1}}},
    \end{align}
    whereby using the lower bound $\binom{K-1+q}{\,q-1\,}
\;\ge\;
\Bigl(\tfrac{K-1+q}{q-1}\Bigr)^{q-1}$ and taking logarithm, we get
%-------------
\begin{align}
           \frac{\delta_{\rm{Majority,1}}}{\delta_{\rm{ChannelComp}}} &\leq  \frac{K\log q}{(q-1)\log\!\Bigl(\tfrac{K+q-1}{q-1}\Bigr)}, \label{eq:upper_major1}\\
         \frac{\delta_{\rm{Majority,2}}}{\delta_{\rm{ChannelComp}}} & \leq \frac{\log{(Kq)}}{(q-1)\log\!\Bigl(\tfrac{K+q-1}{q-1}\Bigr)}, \label{eq:upper_major2}
\end{align}
%-------------
The term in the denominator,  for $K\gg q$, can be written
%-------------
\begin{align}
    \nonumber \log\!\Bigl(\tfrac{K+q-1}{q-1}\Bigr)
&=\log\!\Bigl(\tfrac{K}{q-1}\,(1+O(\tfrac{q}{K}))\Bigr),
\\& =\log\!\Bigl(\tfrac{K}{q-1}\Bigr)+o(1). \label{eq:somthinhghere}
\end{align}
%-------------
Accordingly, the upper bound in \eqref{eq:upper_major1} can be written as
%----------------
\begin{align}
    \frac{\delta_{\rm{Majority,1}}}{\delta_{\rm{ChannelComp}}} &\leq \mathcal{O}\bigg( \frac{K\log q}{(q-1)\log\!\Bigl(\tfrac{K}{q-1}\Bigr)}\bigg), \\
    & =\mathcal{O}\bigg( \frac{K}{(q-1)\log_{q}\!({K})}\bigg)
\end{align}
%----------------
To simplify the upper bound fo the second inequality in \eqref{eq:upper_major2}, we use the equality in \eqref{eq:somthinhghere}, and then we have 
\begin{align}
    \nonumber
    \frac{\log(Kq)}{\log\!\bigl(\tfrac{K+q-1}{q-1}\bigr)}
&=
\frac{\log K + \log q}{\log K - \log(q-1) + o(1)}
\\& =
\frac{1 + \frac{\log q}{\log K}}{1 - \frac{\log(q-1)}{\log K} + o(1)}
=O(1).
\end{align}
It follows that
%-----------
\begin{align}
    \frac{\delta_{\rm Majority,2}}{\delta_{\rm ChannelComp}}
\;\le\;
\frac{1}{q-1}\,O(1)
\;=\;
O\!\bigl(\tfrac{1}{q-1}\bigr),
\end{align}
%-----------
as required. Hence, we can conclude the proof. 

\bibliographystyle{IEEEtran}
\bibliography{IEEEabrv,Ref}

\end{document}

%% file: Figs/Fig_SystemModel.tex
\begin{figure*}
    \centering

\tikzset{every picture/.style={line width=0.75pt}} %set default line width to 0.75pt        

\begin{tikzpicture}[x=0.75pt,y=0.75pt,yscale=-1,xscale=1]

%Shape: Rectangle [id:dp8043797058478284] 

\begin{scope}[shift={(-1.2cm,0)}]

\draw [dashed, color=bazaar]   (383pt,70pt) -- (383pt,165pt) ;
\draw [dashed, color=bazaar]   (433pt,70pt) -- (433pt,165pt) ;

\draw [dashed, color=bazaar]   (485pt,70pt) -- (485pt,165pt) ;

% Rect- D(\cdot)
\draw[fill=palecgray , rounded corners=5pt] (470pt, 130pt) rectangle (440pt, 100pt) {};
\draw (455pt,115pt) node   {{\color{brown(web)}$\mathcal{T}(\cdot)$}};

\draw [-latex]    (420pt,115pt) -- (440pt,115pt) ;

\draw[fill=palecgray , rounded corners=5pt] (520pt, 130pt) rectangle (490pt, 100pt) {};
\draw (505pt,115pt) node   {{\color{brown(web)}$\mathcal{Q}_p(\cdot)$}};

\draw [-latex]    (520pt,115pt) -- (540pt,115pt) ;

\draw[fill=palecgray , rounded corners=5pt] (420pt, 130pt) rectangle (390pt, 100pt) {};
\draw (405pt,115pt) node   {$\mathcal{D}(\cdot)$};
\draw [-latex]    (370pt,115pt) -- (390pt,115pt) ;

\draw (540,80) node {\scriptsize Modulation  Decoding};
\draw (620,80) node {\scriptsize Tabular Mapper };
\draw (680,80) node {\scriptsize Sampling};
%--------------
\draw [-latex]  (310pt,80pt) -- (354pt,110pt) ;
\draw [-latex]  (310pt,150pt) -- (354pt,120pt) ;

\draw (360pt, 115pt) node {\LARGE $\bigoplus$};
\draw (480,105) node {$z$};

\draw [-latex]  (480,85pt) -- (480,105pt) ;

\draw (500,145) node {$r$};
\draw (570,145) node {$\tilde{x}$};

\draw [-latex]  (470pt,115pt) -- (490pt,115pt) ;
\draw (635,143) node {$\hat{f}$};

\draw (710,143) node {$\tilde{f}$};

\draw  [color={rgb, 255:red, 74; green, 144; blue, 226 }][dash pattern={on 0.84pt off 2.51pt}] (505,125) -- (700,125) -- (700,180) -- (505,180) -- cycle ;

\draw (600,115) node {\color{rgb, 255:red, 74; green, 144; blue, 226 } $\mathscr{D}$};

\end{scope}

\draw[-latex]   (200pt,80pt) -- (220pt,80pt) ;

\draw (227pt, 50pt) node { $p_1$};
\draw [-latex]  (227pt,55pt) -- (227pt,75pt) ;
\draw (227pt, 80pt) node { $\bigotimes$};

\draw  [-latex]  (235pt,80pt) -- (255pt,80pt) ;

\draw [-latex]  (227pt,125pt) -- (227pt,145pt) ;
\draw (227pt, 150pt) node { $\bigotimes$};
\draw (227pt, 120pt) node { $p_K$};
\draw [-latex]   (235pt,150pt) -- (255pt,150pt) ;

\draw  [color={rgb, 255:red, 74; green, 144; blue, 226 }][dash pattern={on 0.84pt off 2.51pt}] (85,80) -- (275,80) -- (275,130) -- (85,130) -- cycle ;

% Rect- D(\cdot)
\draw[fill=palecgray , rounded corners=5pt] (200pt, 95pt) rectangle (170pt, 65pt) {};
\draw (185pt,80pt) node   {$\mathcal{E}(\cdot)$};

\draw [-latex]    (150pt,80pt) -- (170pt,80pt) ;

\draw[fill=palecgray , rounded corners=5pt] (150pt, 95pt) rectangle (120pt, 65pt) {};
\draw (135pt,80pt) node   {{\color{brown(web)}$\mathcal{Q}(\cdot)$}};
\draw [-latex]    (100pt,80pt) -- (120pt,80pt) ;

\draw[fill=palecgray , rounded corners=5pt] (100pt, 95pt) rectangle (70pt, 65pt) {};
\draw (85pt,80pt) node   {$\varphi_1(\cdot)$};
\draw [-latex]    (50pt,80pt) -- (70pt,80pt) ;

%--------------
\draw[-latex]    (200pt,150pt) -- (220pt,150pt) ;

\draw[fill=palecgray , rounded corners=5pt] (200pt, 165pt) rectangle (170pt, 135pt) {};
\draw (185pt,150pt) node   {$\mathcal{E}(\cdot)$};

\draw [-latex]    (150pt,150pt) -- (170pt,150pt) ;

\draw[fill=palecgray , rounded corners=5pt] (150pt, 165pt) rectangle (120pt, 135pt) {};
\draw (135pt,150pt) node   {{\color{brown(web)}$\mathcal{Q}(\cdot)$}};
\draw [-latex]    (100pt,150pt) -- (120pt,150pt) ;

\draw[fill=palecgray , rounded corners=5pt] (100pt, 165pt) rectangle (70pt, 135pt) {};
\draw (85pt,150pt) node   {$\varphi_K(\cdot)$};
\draw [-latex]    (50pt,150pt) -- (70pt,150pt) ;

%--------------

\draw  [color={rgb, 255:red, 74; green, 144; blue, 226 }][dash pattern={on 0.84pt off 2.51pt}] (85,175) -- (275,175) -- (275,225) -- (85,225) -- cycle ;

\draw [dashed, color=bazaar]   (114pt,40pt) -- (114pt,175pt) ;
\draw [dashed, color=bazaar]   (164pt,40pt) -- (164pt,175pt) ;

\draw (180,70) node {\color{rgb, 255:red, 74; green, 144; blue, 226 } $\mathscr{E}_1$};

\draw (105,40) node {\scriptsize Source Encoding};
\draw (185,40) node {\scriptsize Quantization};
\draw (270,40) node {\scriptsize Modulation Encoding};

\draw (75,100) node {$s_1$};
\draw (145,100) node {$c_1$};
\draw (210,100) node {$\tilde{c}_1$};
\draw (289,100) node {$x_1$};
\draw (350,100) node {$x_1p_1$};
\draw (400,110) node {$h_1$};

\draw (75,145) node {\Large $\vdots$};
\draw (145,145) node {\Large $\vdots$};
\draw (210,145) node {\Large $\vdots$};
\draw (280,145) node {\Large $\vdots$};
\draw (380,145) node {\Large $\vdots$};

\draw (75,190) node {$s_K$};
\draw (145,190) node {$c_K$};
\draw (210,190) node {$\tilde{c}_K$};
\draw (289,190) node {$x_K$};
\draw (350,190) node {$x_Kp_K$};
\draw (400,200) node {$h_K$};

\draw (180,160) node {\color{rgb, 255:red, 74; green, 144; blue, 226 } $\mathscr{E}_K$};

\end{tikzpicture}

\caption{Block diagram of the communication model. Each node $k$ encodes its value $s_k$ via source encoding $\varphi_k(\cdot)$, quantization $\mathcal{Q}(\cdot)$, and modulation $\mathcal{E}(\cdot)$ to generate $x_k$. The signals are weighted by $p_k$ and transmitted over the MAC with channel coefficients $h_k$. At the CP, the received signal $r$ is processed by a decoder $\mathcal{D}(\cdot)$ and a tabular mapper $\mathcal{T}(\cdot)$ to estimate the function output $\hat{f}$. Finally, the estimated output $\hat{f}$ is quantized (sampled) by $\mathcal{Q}_p(\cdot)$ to yield $\tilde{f}\in \mathcal{Y}_f^{p}$, where $p$ denotes the quantization levels. The encoding operators $\mathscr{E}_k$ and the decoding operator $\mathscr{D}$ enable efficient computation via communication. In this paper, we focus on studying the impact of the quantizer  {\color{brown(web)} $\mathcal{Q}(\cdot)$ } and sampler {\color{brown(web)} $\mathcal{Q}_p(\cdot)$ }, and the tabular operator {\color{brown(web)} $\mathcal{T}(\cdot)$ } which are depicted by color code {\color{brown(web)} brown}. }

    \label{fig:Systemmodel}
\end{figure*}
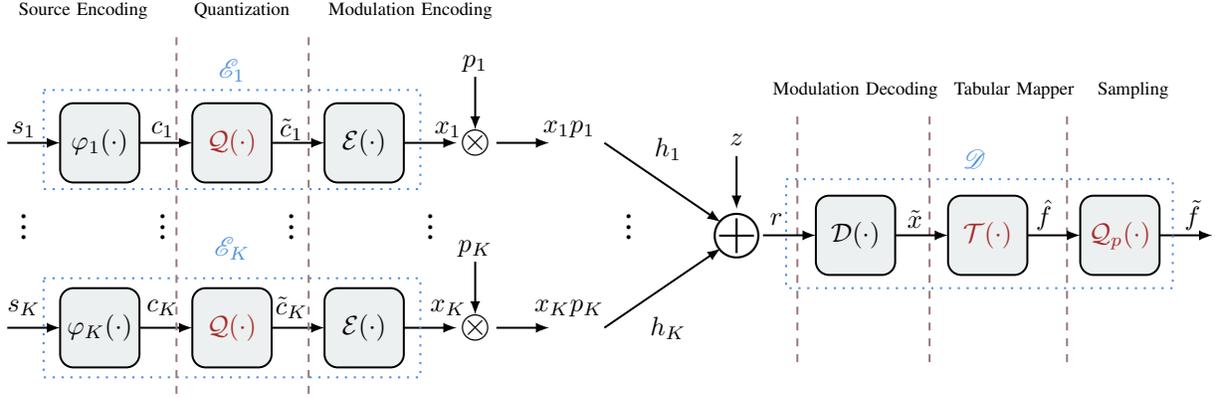

%% file: Figs/Fig_ChannelComp_idea.tex
\begin{figure}
    \centering
  \scalebox{0.8}{  

\tikzset{every picture/.style={line width=0.75pt}} %set default line width to 0.75pt        

\begin{tikzpicture}[x=0.7pt,y=0.7pt,yscale=-1,xscale=1]
%uncomment if require: \path (0,276); %set diagram left start at 0, and has height of 276

%Shape: Ellipse D_f
\draw (140,150) ellipse (1.7cm and 3cm);

%Shape: Ellipse R_f
\draw (405,150) ellipse (1.1cm and 2.8cm);

%Shape: Ellipse \sum_k

\draw[dashed] (270,150) ellipse (0.9cm and 2cm);

\draw (270,85) node    {${r}_{1}$};
% Text Node
\draw (270,100) node   {${r}_{2}$};
\draw (270,110) node     {$\vdots$};
\draw (270,135) node   {${r}_{5}$};
\draw (270,150) node     {$\vdots$};
% Text Node

\draw (270,170) node    {${r}_{n'}$};
\draw (270,185) node    {$\vdots $};
% Text Node

% Text Node
\draw (270,210) node   {${r}_{n}$};

%Straight Lines 
\draw[-latex]    (170,70) -- (260,85) ;

\draw [color=bleudefrance,-latex]   (167,95) -- (260,100) ;

%Straight Lines 
\draw [color=bleudefrance,-latex]   (193,150) -- (260,135) ;

%Straight Lines 
\draw [color=bleudefrance,-latex]   (280,100) -- (385,116) ;
\draw [color=bleudefrance,-latex]   (280,135) -- (385,116) ;
%Straight Lines 
\draw [color=chestnut, -latex]   (285,170) -- (385,175) ;
%Straight Lines 
\draw [color=chestnut,-latex ]   (285,170) -- (390.9,226.39) ;

%Straight Lines [id:da31888424132539517] 
\draw[-latex]    (280,85) -- (385,85) ;

%Straight Lines [id:da817962164978439] 
\draw [color=chestnut,-latex ]   (150,185) -- (260,170) ;
%Straight Lines [id:da612982519144587] 
\draw [color=chestnut,-latex ]   (190,210) -- (260,170) ;

% Text Node
\draw (140,66.4) node  {$0,\dotsc ,0$};
% Text Node
\draw (140,118.4) node    {$\vdots $};
% Text Node
\draw (140,175.4) node     {$\vdots $};
% Text Node

% Text Node
\draw (145,15) node    {$\mathcal{X}^{K}$};
% Text Node
\draw (270,15) node    {$\mathcal{R}_s$};
% Text Node
\draw (405,15) node   {$\mathcal{Y}_g$};
% Text Node
\draw (270,250) node     {$\sum _{k}{x}_{k}$};
% Text Node
\draw  [color=bleudefrance ]  (114,86) -- (167,86) -- (167,106) -- (114,106) -- cycle  ;
\draw (140,95) node  [font=\footnotesize]  {$1,\dotsc ,0$};
% Text Node
\draw  [color=bleudefrance ]  (108,143) -- (193,143) -- (193,163) -- (108,163) -- cycle  ;

\draw (150,155) node   [font=\footnotesize]  {$q-1,\dotsc ,1,0$};
% Text Node
\draw  [color=chestnut  ,draw opacity=1 ]  (97,199) -- (190,199) -- (190,219) -- (97,219) -- cycle  ;
\draw (140,210) node  [font=\footnotesize]  {$q-1,\dotsc ,q-1$};
% Text Node
\draw (405,85) node    {$g^{(1)}$};
% Text Node
\draw (405,112) node    {$g^{(2)}$};
% Text Node
\draw (405,225) node     {$g^{(m)}$};
% Text Node
\draw (405,193.4) node     {$\vdots $};
% Text Node
\draw (405,175) node   {$g^{(m')}$};
% Text Node
\draw (405,136.4) node   {$\vdots $};

% Text Node

\end{tikzpicture}
    }
    \caption{ This figure gives the main idea of Constellation points overlaps~\cite{saeed2023ChannelComp}. It reports the domain $\mathcal{X}^{K}$, the range of the summation $\mathcal{R}_s$ returned by a noise-free channel, and the desired function $\mathcal{Y}_g$ on the left, middle, and right, respectively. For different points (blue lines) in the domain of function yielding constellation points ${r}_2$ and ${r}_5$, distinct received symbols (${r}_i\neq {r}_j$) permit unique association of each $g^{(2)}$ to its symbol. In contrast, inputs shown in red converge to the same received symbol $r_{n'}$ yet yield different function values $g^{(m)}\neq g^{(m')}$, making their assignment impossible without splitting $r_{n'}$ via an appropriate modulation designs.}
    \label{fig:range_vector}
\end{figure}
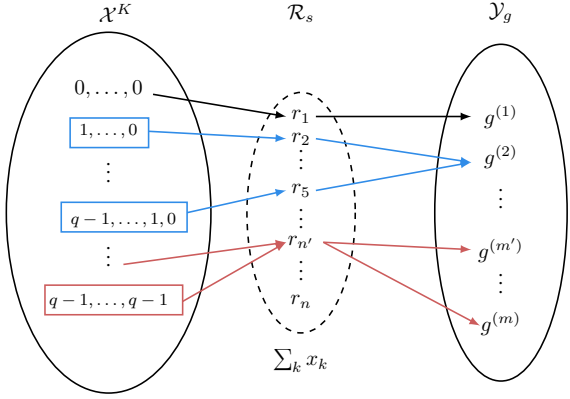

%% file: Figs/Fig_SquareConst.tex
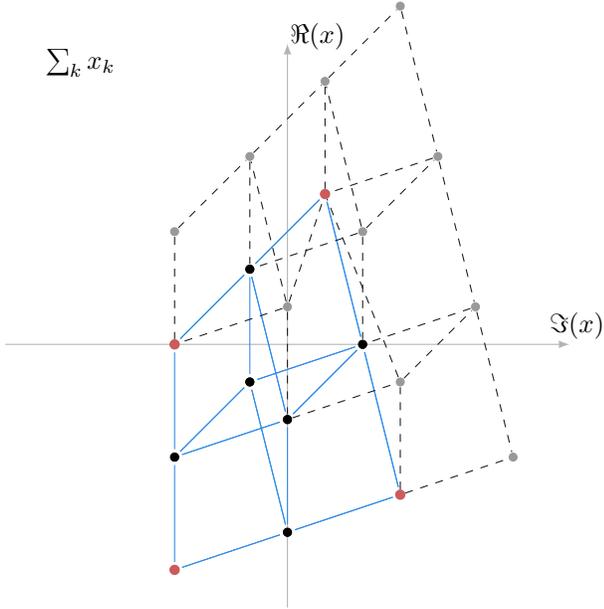
\begin{figure}[t]
    \centering

% Define styles
\tikzset{
    mainEdge/.style={bleudefrance,shorten >=3pt,shorten <=3pt},
    transparentOverlay/.style={ black!90, dashed, shorten >=2pt, shorten <=2pt}
}
\begin{tikzpicture}[scale=0.5]
\begin{scope}[shift={(-3.5cm,-0.5cm)},]
\draw (7.3,14.7) node    {$\Re(x)$};
\draw (14.2,7) node    {$\Im(x)$};
\draw (1,14) node    {$\sum_kx_k$};
\draw[-latex,black!30]    (-1,6.5) -- (14,6.5) ;
\draw[-latex,black!30]    (6.5,-0.5) -- (6.5,14.5) ;
\end{scope}
% Define a general convex quadrilateral (in cyclic order)

\coordinate (s1) at (0,0);
\coordinate (s2) at (3,1);
\coordinate (s3) at (2,5);
\coordinate (s4) at (0,3);

% \coordinate (s1) at (1,0);
% \coordinate (s2) at (4,1);
% \coordinate (s3) at (3,5);
% \coordinate (s4) at (1,3);

% For the transparent overlay we define a translation lattice.
% Here we use the vectors from s1 to s2 and from s1 to s4.
\coordinate (v) at ($ (s2)-(s1) $); % = (4,1)
\coordinate (w) at ($ (s4)-(s1) $); % = (-1,3)
\coordinate (P51) at ($ (s4)+(s4)+(s4)$);
\coordinate (P52) at ($ (s3)+(s4)+(s4)$);
\coordinate (P53) at ($ (s3)+(s3)+(s4)$);
\coordinate (P54) at ($ (s2) +(s3) + (s3)$);
\coordinate (P55) at ($ (s2) +(s2) + (s3)$);
\coordinate (P56) at ($ (s2) +(s2) + (s2)$);

\coordinate (Q1) at ($(s2) + (s2) + (s1)$); 
\coordinate (Q2) at ($(s2)+ (s2) + (s2)$); 
\coordinate (Q3) at ($(s4)+ (s2) + (s2)$); 
\coordinate (Q4) at ($(s2)+ (s2) +(s3)$); 
\coordinate (Q5) at ($(s2)+ (s4)+(s1)$); 
\coordinate (Q6) at ($(s3)+ (s3)+(s1)$);
\coordinate (Q7) at ($(Q6)+ (s4)+(s1)$);
\coordinate (Q8) at ($(s3)+ (s3)+(s3)$);
\coordinate (Q9) at ($(s3)+ (s3)+(s2)$);
\coordinate (Q10) at ($(s3)+ (s2)+(s4)$);
\coordinate (Q11) at ($(s3)+(s4)+(s1)$);
\coordinate (Q12) at ($(s3)+(s2)+(s1)$);
\coordinate (Q13) at ($(s3)+(s3)+(s4)$);
\coordinate (Q14) at ($(s4)+(s4)+(s2)$);
\coordinate (Q15) at ($(s4)+(s4)+(s1)$);
\coordinate (Q16) at ($(s4)+(s4)+(s3)$);
\coordinate (Q17) at ($(s4)+(s1)+(s2)$);
%%%%%%%%%%%%%%%%%%%%%%%%%%%%%%%%%%%%%%%%%%%%%%%%%%%%%%%%%%%%%%
  % Two-User Minkowski Sum Grid (16 points)
  %%%%%%%%%%%%%%%%%%%%%%%%%%%%%%%%%%%%%%%%%%%%%%%%%%%%%%%%%%%%%%
  % For i,j = 1,...,4, define P_ij = s_i + s_j.
  \foreach \i in {1,...,4} {
    \foreach \j in {1,...,4} {
      \coordinate (P\i\j) at ($ (s\i)+(s\j) $);
      % Optionally, label each point:
      % \node[above right,font=\tiny] at (P\i\j) {(\i,\j)};
    }
  }
  % Draw horizontal segments in the base grid
  \foreach \j in {1,...,4} {
    \foreach \i in {1,...,3} {
      \pgfmathtruncatemacro{\iplus}{\i+1}
      \draw[mainEdge] (P\i\j) -- (P\iplus\j);
    }
  }
  % Draw vertical segments in the base grid
  \foreach \i in {1,...,4} {
    \foreach \j in {1,...,3} {
      \pgfmathtruncatemacro{\jplus}{\j+1}
      \draw[mainEdge] (P\i\j) -- (P\i\jplus);
    }
  }
  \draw[mainEdge](P12) -- (P42);
  \draw[mainEdge](P11) -- (P14);
\draw[mainEdge](P14) -- (P44);
\draw[mainEdge](P13) -- (P43);
  
% \draw[mainEdge] ($ (s1)+(s1) $) -- ($ (s2)+(s2) $) -- ($ (s3)+(s3) $) -- ($ (s4)+(s4) $) -- cycle;
    \foreach \i in {1,...,4} {
    \foreach \j in {1,...,4} {
      \filldraw [black] (P\i\j) circle (3pt);
      % Optionally, label each point:
      % \node[above right,font=\tiny] at (P\i\j) {(\i,\j)};
    }
  }
  \filldraw [chestnut] (P11) circle (3.5pt);
  \filldraw [chestnut] (P22) circle (3.5pt);
  \filldraw [chestnut] (P33) circle (3.5pt);
  \filldraw [chestnut] (P44) circle (3.5pt);
  \filldraw [black!40] (P51) circle (3pt);
  \filldraw [black!40] (P52) circle (3pt);
  \filldraw [black!40] (P53) circle (3pt);
  \filldraw [black!40] (P54) circle (3pt);
  \filldraw [black!40] (P55) circle (3pt);
  \filldraw [black!40] (P56) circle (3pt);
   \filldraw [black!40] (Q3) circle (3pt);
     \filldraw [black!40] (Q8) circle (3pt);
     \filldraw [black!40] (Q10) circle (3pt);
     \filldraw [black!40] (Q14) circle (3pt);

  \draw[transparentOverlay] (P51) -- (P52);
\draw[transparentOverlay] (P52) -- (P53);
\draw[transparentOverlay] (P54) -- (P55);
  \draw[transparentOverlay] (P55) -- (P56);

  \draw[transparentOverlay] (P51) -- (P44);
   \draw[transparentOverlay] (P52) -- (P43);
   \draw[transparentOverlay] (P53) -- (P33);
   \draw[transparentOverlay] (P54) -- (P33);
    \draw[transparentOverlay] (P55) -- (P32);
    
     \draw[transparentOverlay] (Q1) -- (Q2);
      \draw[transparentOverlay] (Q1) -- (Q3);
        \draw[transparentOverlay] (Q3) -- (Q4);
          \draw[transparentOverlay] (Q5) -- (Q3);
     \draw[transparentOverlay] (Q3) -- (Q6);
     % \draw[transparentOverlay] (Q6) -- (Q7);
\draw[transparentOverlay] (Q7) -- (Q8);
\draw[transparentOverlay] (Q8) -- (Q9);
\draw[transparentOverlay] (Q10) -- (Q9);
\draw[transparentOverlay] (Q10) -- (Q11);
\draw[transparentOverlay] (Q10) -- (Q12);
\draw[transparentOverlay] (Q10) -- (Q13);
\draw[transparentOverlay] (Q14) -- (Q6);
\draw[transparentOverlay] (Q14) -- (Q15);
\draw[transparentOverlay] (Q14) -- (Q16);
\draw[transparentOverlay] (Q14) -- (Q17);

\end{tikzpicture}
    
    \caption{Quantization for $q=4$  with a sampling order $p=2$. The figure illustrates the discretization of the constellation space into four levels by employing a sampling order $p=2$, highlighting the resulting grid structure derived from the Minkowski sum of the transmitted symbols.}
    \label{fig:SqaureConst}
\end{figure}

%% file: Figs/Fig_table_epsilon.tex
{\color{red!70!black!50}\begin{table}[t]
\centering
\caption{Normalized Sampling Error $\epsilon(p)$ for Common Symmetric Functions}
\begin{tabular}{|l|l|c|c|}
\hline
\textbf{Function} & $g(\bm{s})$ & Lipschitz- $\infty$ norm & \textbf{ $\epsilon(p)$} \\
\hline
{\small Arithmetic Mean} & $\dfrac{1}{K} \sum\limits_{k=1}^{K} s_k$ & $\dfrac{1}{K}$ & $\dfrac{p}{K}$ \\
\hline
{\small Max} & $\max_k\{ s_k \}$ & $1$ & $\dfrac{p}{K}$ \\
\hline
{\small Geometric Mean} & $\left( \prod\limits_{k=1}^{K} s_k \right)^{1/K}$ & $(q-1)$ & $\dfrac{p(q-1)}{K}$ \\
\hline
\end{tabular}\label{tab:epsilon}
\end{table}}

%% file: Figs/Fig_tred_off.tex
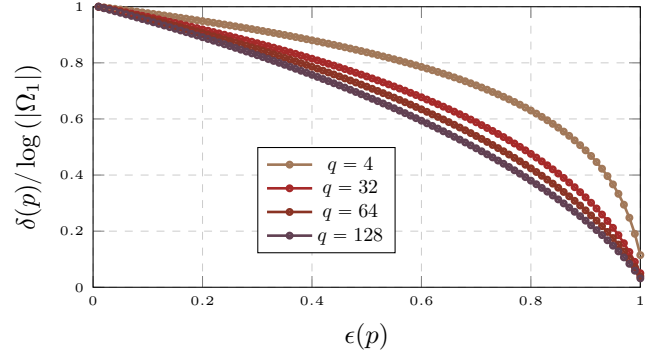
\begin{figure}[t]
  \centering
  \begin{tikzpicture}
    \begin{axis}[
      width=\linewidth,
      height=0.6\linewidth,
      xlabel={$\epsilon(p)$},
      ylabel={$\delta(p)/\log{(|\Omega_1|)}$},
        xmin=0, xmax=1,
        ymin=0, ymax=1,
      grid=major,
      legend pos=north east,
    legend pos=north west,
      legend style={nodes={scale=0.75, transform shape}, at={(0.3,0.5)}}, 
      ticklabel style={font=\tiny},
    ymajorgrids=true,
    xmajorgrids=true,
    grid style=dashed,
    grid=both,
    grid style={line width=.1pt, draw=gray!15},
    major grid style={line width=.2pt,draw=gray!40},
    ]
      \addplot[  mark=o,
      color= chamoisee,
        line width=1pt,
        mark size=1pt] table [x=delta, y=beta4] {Data/tred_off.dat};
      \addlegendentry{$q=4$}

      \addplot[  mark=o,
      color= brown(web),
        line width=1pt,
        mark size=1pt] table [x=delta, y=beta32] {Data/tred_off.dat};
      \addlegendentry{$q=32$}

      \addplot[  mark=o,
      color= burntumber,
        line width=1pt,
        mark size=1pt] table [x=delta, y=beta64] {Data/tred_off.dat};
      \addlegendentry{$q=64$}
      
          \addplot[  mark=o,
      color= eggplant,
        line width=1pt,
        mark size=1pt] table [x=delta, y=beta128] {Data/tred_off.dat};
      \addlegendentry{$q=128$}
    \end{axis}
  \end{tikzpicture}
  \caption{Trade‐off curves between the worst‐case approximation error $\epsilon(p)$ and the enumeration complexity $\delta(p)$ for sampling orders $q\in\{4,32,64\}$, with $K=100$ and Lipschitz constant $L=1$.  Allowing larger errors yields dramatic reductions in combinatorial complexity.}
  \label{fig:tradeoff}
\end{figure}

%% file: Figs/Fig_Qunatization_plot.tex
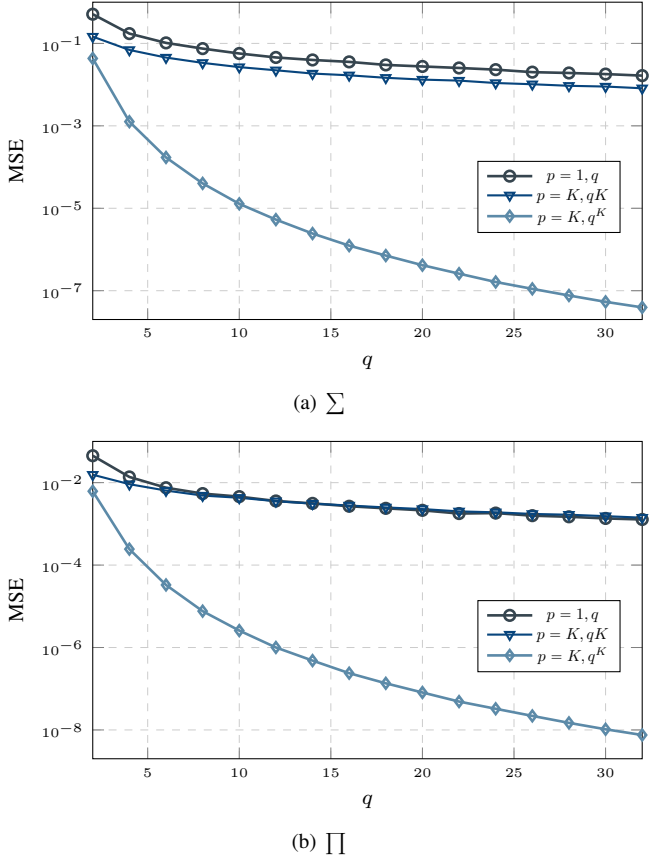
\begin{figure}
    \centering
\subfigure[$\sum$]{
\begin{tikzpicture}
  \begin{axis}[
      xlabel={$q$},
      ylabel={MSE},
    label style={font=\footnotesize},
        width=0.49\textwidth,
        height=0.32\textwidth,,
        xmin=2, xmax=32,
        ymin=2e-8, ymax=1,
      ymode=log,
      legend pos=north west,
      legend style={nodes={scale=0.6, transform shape}, at={(0.7,0.5)}}, 
      ticklabel style={font=\tiny},
    ymajorgrids=true,
    xmajorgrids=true,
    grid style=dashed,
    grid=both,
    grid style={line width=.1pt, draw=gray!15},
    major grid style={line width=.2pt,draw=gray!40},
  ]
    % Plot for f_sum: input quantization MSE
    \addplot[
      mark=o,
      color= charcoal,
        line width=1pt,
        mark size=2pt,
    ] table [x index=0, y index=1, col sep=space] {Data/errors_k5.dat};

    \addplot[
      mark=triangle,
      color=darkcerulean,
        mark options={rotate=180},
        line width=1pt,
        mark size=2pt,
      thick,
    ] table [x index=0, y index=3, col sep=space] {Data/errors_k5.dat};
    \addplot[mark=diamond,
      mark options={rotate=180},
      color= airforceblue,
        line width=1pt,
        mark size=2pt,
    ] table [x index=0, y index=2, col sep=space] {Data/errors_k5.dat};
    % Plot for f_prod: input quantization MSE

    % Plot for f_sum: input & output quantization MSE
    % \addplot[
    %   mark=star,
    %   color=chestnut,
    %   thick,
    %     mark options={rotate=180},
    %     line width=1pt,
    %     mark size=2pt,
    % ] table [x index=0, y index=2, col sep=space] {Data/Quantize.dat};
    % \addlegendentry{Input Quantization MSE (prod)}
    
    % Plot for f_prod: input & output quantization MSE
    % \addplot[
    %   mark=diamond,
    %     mark options={rotate=180},
    %     line width=1pt,
    %     mark size=2pt,
    %   color=cadmiumorange,
    %   thick,
    % ] table [x index=0, y index=4, col sep=space] {Data/Quantize.dat};
    % \addlegendentry{Input \& Output Quantization MSE (prod)}
    % \legend{ $E_1 \prod$,  $E_1 \sum$,  $E_2 \prod$, $E_2\sum$};
    \legend{ {$p=1, q$}, {$p=K, qK$},{$p=K, q^{K}$}};
  \end{axis}
\end{tikzpicture}}

\subfigure[$\prod$]{
\begin{tikzpicture}
  \begin{axis}[
      xlabel={$q$},
      ylabel={MSE},
    label style={font=\footnotesize},
        width=0.49\textwidth,
        height=0.32\textwidth,,
        xmin=2, xmax=32,
        ymin=2e-9, ymax=1e-1,
      ymode=log,
      legend pos=north west,
      legend style={nodes={scale=0.6, transform shape}, at={(0.7,0.5)}}, 
      ticklabel style={font=\tiny},
    ymajorgrids=true,
    xmajorgrids=true,
    grid style=dashed,
    grid=both,
    grid style={line width=.1pt, draw=gray!15},
    major grid style={line width=.2pt,draw=gray!40},
  ]
    % Plot for f_sum: input quantization MSE
    \addplot[
      mark=o,
      color= charcoal,
        line width=1pt,
        mark size=2pt,
    ] table [x index=0, y index=4, col sep=space] {Data/errors_k5.dat};
    \addplot[
      mark=triangle,
      color=darkcerulean,
        mark options={rotate=180},
        line width=1pt,
        mark size=2pt,
      thick,
    ] table [x index=0, y index=6, col sep=space] {Data/errors_k5.dat};
    \addplot[mark=diamond,
      mark options={rotate=180},
      color= airforceblue,
        line width=1pt,
        mark size=2pt,
    ] table [x index=0, y index=5, col sep=space] {Data/errors_k5.dat};
       \legend{ {$p=1, q$}, {$p=K, qK$},{$p=K, q^{K}$}};;
  \end{axis}
\end{tikzpicture}}
    
    \caption{MSE versus input quantization level $q\in\{2,4,\dots,32\}$ for the sum (a) and product (b) functions over $K=5$ inputs, comparing three schemes: $p=1$ with order $q$, $p=K$ with order $qK$, and $p=K$ with order $q^K$. The MSE is computed by Monte Carlo simulation ($2\times 10^3$ samples in $[0,1]^5$) and plotted on a logarithmic scale to highlight the error decay achieved by increasing the sampling/modulation level.}
    \label{fig:quantize}
\end{figure}

%% file: Figs/Fig_TriangleCont.tex
\begin{figure}[t]
    \centering

\begin{tikzpicture}[scale=1]
    % Parameters for edge appearance
    \def\shortenAmt{4pt}    % gap at each end of an edge

   \tikzset{edgeStyle/.style={bleudefrance}}
    % Parameters for the transmitted (triangle) constellation
    \def\L{2}    % s2 = (L,0)
    \def\p{1}    % s3 = (p,q)
    \def\q{1.8}  % s3 = (p,q)
    % Define received points (sums of transmitted points)
    % Vertices of the outer triangle:
    \coordinate (P1) at (0,0);         % s1+s1
    \coordinate (P3) at (2*\L,0);       % s2+s2
    \coordinate (P6) at (2*\p,2*\q);     % s3+s3
    
    % Midpoints of the sides:
    \coordinate (P2) at (\L,0);         % s1+s2 (midpoint of P1 and P3)
    \coordinate (P4) at (\p,\q);         % s1+s3 (midpoint of P1 and P6)
    \coordinate (P5) at (\L+\p,\q);      % s2+s3 (midpoint of P3 and P6)

    % Draw the outer triangle (vertices: P1, P3, P6)
    \draw[edgeStyle, shorten >=\shortenAmt, shorten <=\shortenAmt] (P1) -- (P2);
    \draw[edgeStyle, shorten >=\shortenAmt, shorten <=\shortenAmt] (P2) -- (P3);
    \draw[edgeStyle, shorten >=\shortenAmt, shorten <=\shortenAmt] (P3) -- (P5);
    \draw[edgeStyle, shorten >=\shortenAmt, shorten <=\shortenAmt] (P5) -- (P6);
    \draw[edgeStyle, shorten >=\shortenAmt, shorten <=\shortenAmt] (P1) -- (P4);
    \draw[edgeStyle, shorten >=\shortenAmt, shorten <=\shortenAmt] (P4) -- (P6);
    
    % Draw the medial (inner) triangle connecting the midpoints: P2, P5, P4.
    \draw[edgeStyle, shorten >=\shortenAmt, shorten <=\shortenAmt] (P2) -- (P5);
    \draw[edgeStyle, shorten >=\shortenAmt, shorten <=\shortenAmt] (P5) -- (P4);
    \draw[edgeStyle, shorten >=\shortenAmt, shorten <=\shortenAmt] (P4) -- (P2);

    \draw[dashed, color={chestnut},thick,shorten <={15pt}] (P1) -- (P5);
    \draw[dashed, color={chestnut},thick,shorten >={4pt},shorten <={4pt}] (P4) -- (P3);
      \draw[dashed, color={chestnut},thick,shorten >={4pt},shorten <={4pt}] (P6) -- (P2);
        % Plot and label the six points
    \foreach \pt/\lbl in {P1/1, P2/2, P3/3, P4/4, P5/5, P6/6} {
      \filldraw [black] (\pt) circle (2pt);
      \node[above right, font=\footnotesize] at (\pt) {$~r_{\lbl}$};
    }
    
    % Optionally, draw extra segments from vertices to the opposite midpoints
    % to emphasize the four sub-triangles:

    % \draw[\edgeStyle, shorten >=\shortenAmt, shorten <=\shortenAmt] (P3) -- (P4);
\end{tikzpicture}
    \caption{Illustration of the induced constellation diagram of \( \sum_{k}x_k \) for \( q=3, K=2 \). The MAC produces six constellation points \(\{r_1, r_2, \dots, r_6\}\), where the constraints in \eqref{eq:d_x_relax} enforce separation along the solid blue edges, preventing overlaps among the primary constellation points. The red dashed lines indicate unconstrained edges, which may lead to occasional overlaps between the pairs \(\{r_1, r_5\}\), \(\{r_3, r_4\}\), and \(\{r_2, r_6\}\). }
    \label{fig:Trinagle}
\end{figure}
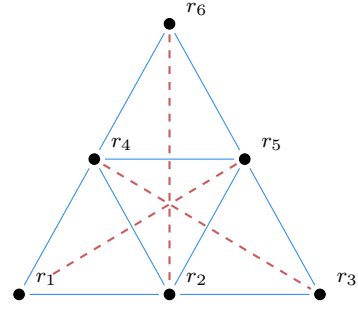

%% file: Figs/Fig_majority.tex
\begin{figure}
    \centering

\scalebox{0.8}{
\tikzset{every picture/.style={line width=0.75pt}} %set default line width to 0.75pt        

\begin{tikzpicture}[x=0.75pt,y=0.75pt,yscale=-0.65,xscale=0.65]
%uncomment if require: \path (0,357); %set diagram left start at 0, and has height of 357

%Straight Lines [id:da5091762878090007] 
\draw    (14.83,100) -- (185.25,99.98) ;
%Straight Lines [id:da11559396899826946] 
\draw    (100.37,177.52) -- (100.75,22) ;

%Straight Lines [id:da5091762878090007] 
\draw    (14.83,100.4) -- (185.25,99.98) ;
%Straight Lines [id:da11559396899826946] 
\draw    (100.37,177.52) -- (100.75,22) ;
%Shape: Ellipse [id:dp897317385969068] 
\draw [dash pattern={on 4.5pt off 4.5pt}] (101,100) node{}  circle  (51);
\draw [fill=black!] (101,50) node{}  circle  (3);
\draw [fill=black!] (101,150) node{}  circle  (3);
\draw [fill=black!] (50,100) node{}  circle  (3);
\draw [fill=black!] (151,100) node{}  circle  (3);

% \draw (113,37) node  {$\vec{j}$};

% \draw (113,205) node  {$\vec{j}$};

\draw [dash pattern={on 4.5pt off 4.5pt}] (101,270) node{}  circle  (51);
\draw [fill=black!] (101,220) node{}  circle  (3);
\draw [fill=black!] (101,320) node{}  circle  (3);
\draw [fill=black!] (50,270) node{}  circle  (3);
\draw [fill=black!] (151,270) node{}  circle  (3);
\draw    (15,270) -- (185, 270) ;

\draw    (390,180) -- (610,180) ;
%Straight Lines [id:da1168551736197786] 
\draw    (500,289.98) -- (500,68.22) ;

\draw [color=chestnut, fill=chestnut] (400,180) node{}  circle  (3);
\draw [color=chestnut, fill=chestnut] (600,180) node{}  circle  (3);
\draw [color=black!50, fill=black!50] (500,180) node{}  circle  (3);
\draw [color=chestnut, fill=chestnut] (500,80) node{}  circle  (3);
\draw [color=chestnut, fill=chestnut] (500,280) node{}  circle  (3);
\draw [color=black!50, fill=black!50] (440,240) node{}  circle  (3);
\draw [color=black!50, fill=black!50] (560,240) node{}  circle  (3);
\draw [color=black!50, fill=black!50] (440,120) node{}  circle  (3);
\draw [color=black!50, fill=black!50] (560,120) node{}  circle  (3);

%Shape: Ellipse [id:dp18451122444088797] 

\draw  [dash pattern={on 0.84pt off 2.51pt}] (565.47,204.58) -- (528.75,243.99) -- (475.43,245.39) -- (436.76,207.96) -- (435.38,153.63) -- (472.11,114.22) -- (525.42,112.82) -- (564.09,150.25) -- cycle ;
%Straight Lines [id:da5898095061789436] 
\draw  [dash pattern={on 0.84pt off 2.51pt}]  (564.09,150.24) -- (615.42,136.06) ;
%Straight Lines [id:da6863816710089473] 
\draw  [dash pattern={on 0.84pt off 2.51pt}]  (525.42,112.81) -- (544.11,60) ;
%Straight Lines [id:da8432654627568398] 
\draw  [dash pattern={on 0.84pt off 2.51pt}]  (472.11,114.22) -- (448.76,63.38) ;
%Straight Lines [id:da9261062937946722] 
\draw  [dash pattern={on 0.84pt off 2.51pt}]  (435.38,153.63) -- (383.25,141.97) ;
%Straight Lines [id:da5570343481652726] 
\draw  [dash pattern={on 0.84pt off 2.51pt}]  (460.36,300) -- (475.43,245.39) ;
%Straight Lines [id:da09866909150182601] 
\draw  [dash pattern={on 0.84pt off 2.51pt}]  (547.43,295.77) -- (528.75,243.99) ;
%Straight Lines [id:da8251634233405793] 
\draw  [dash pattern={on 0.84pt off 2.51pt}]  (616.25,220.56) -- (565.47,204.58) ;
%Straight Lines [id:da9033960418186178] 
\draw  [dash pattern={on 0.84pt off 2.51pt}]  (384.08,223.1) -- (436.76,207.97) ;

\draw    (100.37,345.52) -- (100.75,190) ;

\draw (280,180) node  {\Large $\bigoplus$};

%Straight Lines [id:da2888765950132379] 
\draw[-latex]    (188.25,135) -- (261.28,168.74) ;
% \draw [shift={(264,170)}, rotate = 204.8] [fill={rgb, 255:red, 0; green, 0; blue, 0 }  ][line width=0.08]  [draw opacity=0] (8.93,-4.29) -- (0,0) -- (8.93,4.29) -- cycle    ;
%Straight Lines [id:da14786363787588608] 
\draw[-latex]    (196.25,242)  -- (258.63,193.83) ;
% \draw [shift={(261,192)}, rotate = 142.32] [fill={rgb, 255:red, 0; green, 0; blue, 0 }  ][line width=0.08]  [draw opacity=0] (8.93,-4.29) -- (0,0) -- (8.93,4.29) -- cycle    ;
%Straight Lines [id:da7011524793954549] 
\draw [-latex]   (300,180) -- (345,180) ;
% \draw [shift={(340.25,182)}, rotate = 180] [fill={rgb, 255:red, 0; green, 0; blue, 0 }  ][line width=0.08]  [draw opacity=0] (8.93,-4.29) -- (0,0) -- (8.93,4.29) -- cycle    ;

%-----------------

% Text Node
\draw (160, 80) node  {$0$};
% Text Node
\draw (40,80) node {$3$};
% Text Node
\draw (110,140) node  {$2$};
% Text Node
\draw (110,30) node  {$1$};
% Text Node
%-----------------
%---------------
\draw (160, 250) node   {$0$};
% Text Node
\draw (40,250) node {$3$};
% Text Node
\draw (110,300) node   {$2$};
% Text Node
\draw (110,200) node    {$1$};
% Text Node

% Text Node
\draw (250,137) node [anchor=north west][inner sep=0.75pt]  [font=\footnotesize] [align=left] {Over-the-Air};

\draw (440,9.4) node [anchor=north west][inner sep=0.75pt]  [color={rgb, 255:red, 8; green, 80; blue, 177 }  ,opacity=1 ]  {$f_{1}( x_{1} ,x_{2}) =x_{1} x_{2}$};
% Text Node
\draw (619.76,168.6) node [anchor=north west][inner sep=0.75pt]  [color={rgb, 255:red, 8; green, 80; blue, 177 }  ,opacity=1 ] [align=left] {$\displaystyle 0$};
% Text Node
\draw (565.39,250.55) node [anchor=north west][inner sep=0.75pt]  [color=black!50]  [align=left] {$\displaystyle 0$};
% Text Node
\draw (558.39,90.55) node [anchor=north west][inner sep=0.75pt] [color=black!50]  [align=left] {$\displaystyle 0$};
% Text Node
\draw (505,158.6) node [anchor=north west][inner sep=0.75pt] [color=black!50]  [align=left] {$\displaystyle 0$};
% Text Node
\draw (493.39,42.55) node [anchor=north west][inner sep=0.75pt]  [color={rgb, 255:red, 8; green, 80; blue, 177 }  ,opacity=1 ] [align=left] {$\displaystyle 1$};
% Text Node
\draw (420.39,91.55) node [anchor=north west][inner sep=0.75pt]  [color=black!50]  [align=left] {$\displaystyle 3$};
% Text Node
\draw (370.39,167.55) node [anchor=north west][inner sep=0.75pt]  [color={rgb, 255:red, 8; green, 80; blue, 177 } ] [align=left] {$9$};
% Text Node
\draw (416.39,238.55) node [anchor=north west][inner sep=0.75pt]  [color=black!50] [align=left] {$ 6$};
% Text Node
\draw (494.39,300.55) node [anchor=north west][inner sep=0.75pt]  [color={rgb, 255:red, 8; green, 80; blue, 177 }  ,opacity=1 ] [align=left] {$\displaystyle 4$};
% Text Node
\draw (485.39,158.55) node [anchor=north west][inner sep=0.75pt]  [color=black!50]  [align=left] {$\displaystyle 2$};

\end{tikzpicture}
}
    
\caption{Illustration of the sampling scheme for \(q=4\) computing the product function \(f(s_{1},s_{2}) = s_{1}s_{2}\), with \(s_{1},s_{2}\in\{0,1,2,3\}\). Only the corner points of the superimposed constellation are employed; overlaps at the centre between output symbols “2” and “0” are ignored, simplifying the computation at the expense of reduced accuracy.}

    \label{fig:majority}
\end{figure}
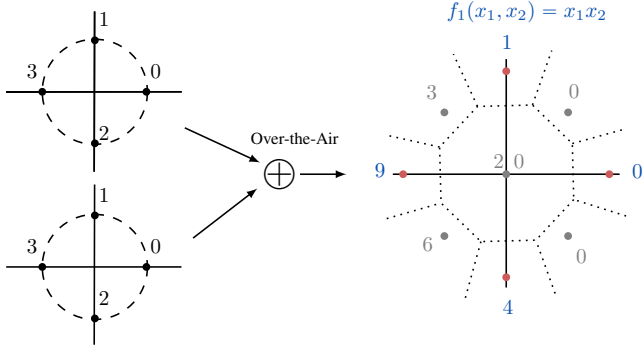

%% file: Figs/Fig_table_compare.tex
\begin{table*}[!t]
\centering
\caption{Comparison of Sampling Schemes for Different Levels \(p\) and Alphabets \(q\)}
\label{tab:sampling-comparison-full}
\begin{tabular}{@{}c c c c p{5.5cm}@{}}
\toprule
\(p\) & \(q\) & Desired function \(g\) & Operator \(T_{p}\) & Method \\ 
\midrule
1 
  & any \(q\) 
  & symmetry \(g\) 
  & \(T_{1}=\mathcal{I}\) 
  & Full enumeration over \(\mathcal{X}^K\) (ChannelComp \cite{saeed2023ChannelComp}) \\[6pt]

\(1<p<K\) 
  & any \(q\) 
  & symmetry \(g\) 
  &  \(\mathcal{T}_{p}\) 
  & Trade-off between complexity and accuracy \\[6pt]

\multirow{2}{*}{\(K\)} 
  & $2$ 
  & \(g(x)=\frac{1}{K}\sum_{i=1}^K x_i\) 
  & \(T_{K}(g)=\operatorname{sign}\!\bigl(g\bigr)\) 
  & OBDA~\cite{zhu2020one}, for $\mathcal{X}=\{-1,1\}$ \\ 

  & \(>2\) 
  & \(g(x)=Q\!\bigl(\tfrac1K\sum_{i=1}^K x_i\bigr)\) 
  & \(T_{\Omega_K}(x)=\arg\min_{a\in\mathcal X}\Bigl|\tfrac1K\sum_{i=1}^K x_i - a\Bigr|\) 
  & Balanced‐vote in \cite{csahin2023over}, for \(\mathcal A\) be balanced numeral numbers. Also, voting happens over constellation points, not frequency.  \\
\bottomrule
\end{tabular}\label{tab:compraison}
\end{table*}

%% file: Figs/Fig_Num1.tex
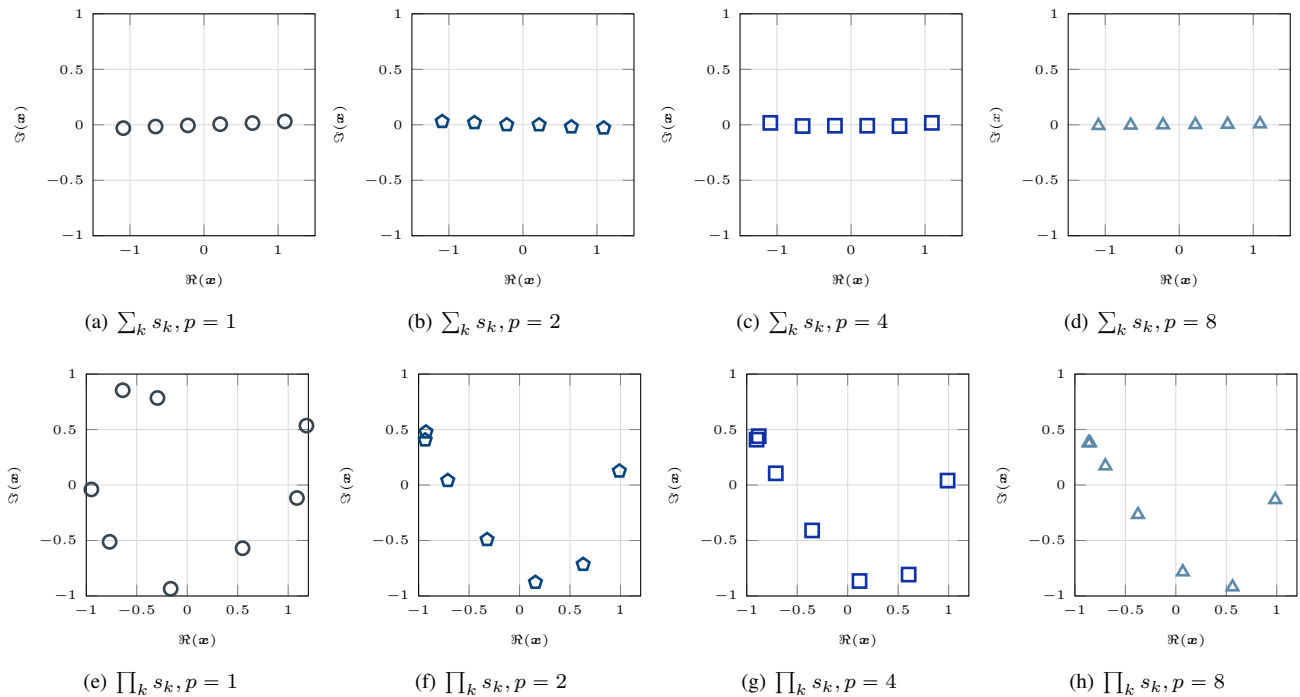
\begin{figure*}[t]
  \centering
  % --- Sum (row 1) ---
  \subfigure[$\sum_k s_k, p=1$]{%
    \begin{tikzpicture}
    \begin{axis}[
      xlabel={$\Re(\bm{x})$}, ylabel={$\Im(\bm{x})$},
      label style={font=\tiny}, tick label style={font=\tiny},
      width=0.25\textwidth, height=0.25\textwidth,
      xmin=-1.5, xmax=1.5, ymin=-1, ymax=1,
      grid=both, grid style={line width=.1pt, draw=gray!10},
      major grid style={line width=.2pt,draw=gray!30},
    ]
      \addplot[
        color=charcoal, line width=1pt, mark=o, only marks, mark size=2.5pt
      ] table[col sep=space, x=real,y=imag] {Data/sum_const_full.dat};
    \end{axis}
    \end{tikzpicture}}
  \subfigure[$\sum_k s_k, p=2$]{%
    \begin{tikzpicture}
    \begin{axis}[
      xlabel={$\Re(\bm{x})$}, ylabel={$\Im(\bm{x})$},
      label style={font=\tiny}, tick label style={font=\tiny},
      width=0.25\textwidth, height=0.25\textwidth,
      xmin=-1.5, xmax=1.5, ymin=-1, ymax=1,
      grid=both, grid style={line width=.1pt, draw=gray!10},
      major grid style={line width=.2pt,draw=gray!30},
    ]
      \addplot[
        color=darkcerulean,line width=1pt, 
        mark=pentagon, only marks, mark size=2.5pt
      ] table[col sep=space, x=real,y=imag] {Data/sum_const_p1.dat};
    \end{axis}
    \end{tikzpicture}
  }
  \subfigure[$\sum_k s_k, p=4$]{%
    \begin{tikzpicture}
    \begin{axis}[
      xlabel={$\Re(\bm{x})$}, ylabel={$\Im(\bm{x})$},
      label style={font=\tiny}, tick label style={font=\tiny},
      width=0.25\textwidth, height=0.25\textwidth,
      xmin=-1.5, xmax=1.5, ymin=-1, ymax=1,
      grid=both, grid style={line width=.1pt, draw=gray!10},
      major grid style={line width=.2pt,draw=gray!30},
    ]
      \addplot[
        color=egyptianblue,line width=1pt, mark=square, only marks, mark size=2.5pt
      ] table[col sep=space, x=real,y=imag] {Data/sum_const_p2.dat};
    \end{axis}
    \end{tikzpicture}
  }
  \subfigure[$\sum_k s_k, p=8$]{%
    \begin{tikzpicture}
    \begin{axis}[
      xlabel={$\Re(\bm{x})$}, ylabel={$\Im({x})$},
      label style={font=\tiny}, tick label style={font=\tiny},
      width=0.25\textwidth, height=0.25\textwidth,
      xmin=-1.5, xmax=1.5, ymin=-1, ymax=1,
      grid=both, grid style={line width=.1pt, draw=gray!10},
      major grid style={line width=.2pt,draw=gray!30},
    ]
      \addplot[
        color=airforceblue,line width=1pt, mark=triangle, only marks, mark size=2.5pt
      ] table[col sep=space, x=real,y=imag] {Data/sum_const_p4.dat};
    \end{axis}
    \end{tikzpicture}
  }

  % --- Prod (row 2) ---
  \subfigure[$\prod_k s_k, p=1$]{%
    \begin{tikzpicture}
    \begin{axis}[
      xlabel={$\Re(\bm{x})$}, ylabel={$\Im(\bm{x})$},
      label style={font=\tiny}, tick label style={font=\tiny},
      width=0.25\textwidth, height=0.25\textwidth,
      xmin=-1, xmax=1.2, ymin=-1, ymax=1,
      grid=both, grid style={line width=.1pt, draw=gray!10},
      major grid style={line width=.2pt,draw=gray!30},
    ]
      \addplot[
        color=charcoal,line width=1pt, mark=o, only marks, mark size=2.5pt
      ] table[col sep=space, x=real,y=imag] {Data/const_full.dat};
    \end{axis}
    \end{tikzpicture}
  }
  \subfigure[$\prod_k s_k, p=2$]{%
    \begin{tikzpicture}
    \begin{axis}[
      xlabel={$\Re(\bm{x})$}, ylabel={$\Im(\bm{x})$},
      label style={font=\tiny}, tick label style={font=\tiny},
      width=0.25\textwidth, height=0.25\textwidth,
      xmin=-1, xmax=1.2, ymin=-1, ymax=1,
      grid=both, grid style={line width=.1pt, draw=gray!10},
      major grid style={line width=.2pt,draw=gray!30},
    ]
      \addplot[
        color=darkcerulean,line width=1pt, mark=pentagon, only marks, mark size=2.5pt
      ] table[col sep=space, x=real,y=imag] {Data/const_p1.dat};
    \end{axis}
    \end{tikzpicture}
  }
  \subfigure[$\prod_k s_k, p=4$]{%
    \begin{tikzpicture}
    \begin{axis}[
      xlabel={$\Re(\bm{x})$}, ylabel={$\Im(\bm{x})$},
      label style={font=\tiny}, tick label style={font=\tiny},
      width=0.25\textwidth, height=0.25\textwidth,
      xmin=-1, xmax=1.2, ymin=-1, ymax=1,
      grid=both, grid style={line width=.1pt, draw=gray!10},
      major grid style={line width=.2pt,draw=gray!30},
    ]
      \addplot[
        color=egyptianblue, line width=1pt,mark=square, only marks, mark size=2.5pt
      ] table[col sep=space, x=real,y=imag] {Data/const_p2.dat};
    \end{axis}
    \end{tikzpicture}
  }
  \subfigure[$\prod_k s_k, p=8$]{%
    \begin{tikzpicture}
    \begin{axis}[
      xlabel={$\Re(\bm{x})$}, ylabel={$\Im(\bm{x})$},
      label style={font=\tiny}, tick label style={font=\tiny},
      width=0.25\textwidth, height=0.25\textwidth,
      xmin=-1, xmax=1.2, ymin=-1, ymax=1,
      grid=both, grid style={line width=.1pt, draw=gray!10},
      major grid style={line width=.2pt,draw=gray!30},
    ]
      \addplot[
        color=airforceblue,line width=1pt, mark=triangle, only marks, mark size=2.5pt
      ] table[col sep=space, x=real,y=imag] {Data/const_p4.dat};
    \end{axis}
    \end{tikzpicture}
  }

  \caption{Constellation diagrams under full‐enumeration (AWGN) and pyramid sampling orders $p\in\{1,2,4,8\}$ for the sum and product mappings.  Each row corresponds to one function, and each column to one design method.}
  \label{fig:constellation-pyramid}
\end{figure*}

%% file: Figs/Fig_Num2.tex
\begin{figure*}[t]
  \centering
  \subfigure[$(\prod_ks_k)^{1/K}$]{%
    \begin{tikzpicture}
    \begin{axis}[
      xlabel={SNR(dB)},
      ylabel={MSE},
      label style={font=\scriptsize},
      tick label style={font=\scriptsize},
      width=0.48\textwidth,
      height=6.5cm,
      xmin=0, xmax=20,
      ymode=log,
       minor tick num=5,
      legend cell align={left},
      legend style={nodes={scale=0.55, transform shape}, at={(0.98,0.5)}},
      ymajorgrids=true,
      xmajorgrids=true,
      grid=both,
      grid style={line width=.1pt, draw=gray!15},
      major grid style={line width=.2pt, draw=gray!40},
      every axis plot/.append style={line width=1pt}
    ]
    \addplot[color=chamoisee,   mark=square,          mark size=2pt,line width=1pt]
      table[x=SNR,y=QAM16] {Data/prod.dat};

    \addplot[color=amethyst,       mark=triangle,   mark options = {rotate = 180},   mark size=2pt,line width=1pt]
      table[x=SNR,y=PAM16] {Data/prod.dat};

    \addplot[color=airforceblue, mark=+, mark size=2pt,line width=1pt]
      table[x=SNR,y=Hex16] {Data/prod.dat};

    \addplot[color=burntumber, mark=pentagon,   mark size=2pt,line width=1pt]
      table[x=SNR,y=QAM32] {Data/prod.dat};

    \addplot[color=brown(web),      mark=triangle,   mark options = {rotate = 90}, mark size=2pt,line width=1pt]
      table[x=SNR,y=PAM32] {Data/prod.dat};

    \addplot[color=darkcerulean, mark=x,         mark size=2pt,line width=1pt]
      table[x=SNR,y=Hex32] {Data/prod.dat};

    \addplot[color=eggplant, mark=o,           mark size=2pt,line width=1pt]
      table[x=SNR,y=QAM64] {Data/prod.dat};

    \addplot[color=darklavender,     mark=triangle,         mark size=2pt,line width=1pt]
      table[x=SNR,y=PAM64] {Data/prod.dat};

    \addplot[color=charcoal, mark=star,  mark size=2pt,line width=1pt]
      table[x=SNR,y=Hex64] {Data/prod.dat};%\addlegendentry{Hex64}
    \end{axis}
    \end{tikzpicture}
  }\subfigure[$\max_ks_k$]{%
    \begin{tikzpicture}
    \begin{axis}[
      xlabel={SNR  (dB)},
      ylabel={MSE},
      label style={font=\scriptsize},
      tick label style={font=\scriptsize},
      width=0.48\textwidth,
      height=6.5cm,
      xmin=0, xmax=20,
      ymode=log,
      legend cell align={left},
      legend style={nodes={scale=0.55, transform shape}, at={(0.98,0.98)}},
      ymajorgrids=true,
      xmajorgrids=true,
      grid=both,
      grid style={line width=.1pt, draw=gray!15},
      major grid style={line width=.2pt, draw=gray!40},
      every axis plot/.append style={line width=1pt}
    ]
      \addplot[color=chamoisee,   mark=square,         mark size=2pt,line width=1pt] table[x=SNR,y=QAM16] {Data/max.dat}; \addlegendentry{QAM16}
      \addplot[color=amethyst,       mark=triangle,   mark options = {rotate = 180},  mark size=2pt,line width=1pt] table[x=SNR,y=PAM16] {Data/max.dat}; \addlegendentry{PAM16}
      \addplot[color=airforceblue, mark=+, mark size=2pt,line width=1pt] table[x=SNR,y=Hex16] {Data/max.dat}; \addlegendentry{Hex16}
      \addplot[color=burntumber, mark=pentagon,  mark size=2pt,line width=1pt] table[x=SNR,y=QAM32] {Data/max.dat}; \addlegendentry{QAM32}
      \addplot[color=brown(web),  line width=1pt,    mark=triangle,   mark options = {rotate = 90}, mark size=2pt] table[x=SNR,y=PAM32] {Data/max.dat}; \addlegendentry{PAM32}
      \addplot[color=darkcerulean, mark=x,         mark size=2pt,line width=1pt] table[x=SNR,y=Hex32] {Data/max.dat}; \addlegendentry{Hex32}
      \addplot[color=eggplant, mark=o,         mark size=2pt,line width=1pt] table[x=SNR,y=QAM64] {Data/max.dat}; \addlegendentry{QAM64}
      \addplot[color=darklavender,     mark=triangle,        mark size=2pt,line width=1pt] table[x=SNR,y=PAM64] {Data/max.dat}; \addlegendentry{PAM64}
      \addplot[color=charcoal, mark=star,      mark size=2pt,line width=1pt] table[x=SNR,y=Hex64] {Data/max.dat}; \addlegendentry{Hex64}
    \end{axis}
    \end{tikzpicture}
  }

  \caption{MSE versus per‐node SNR for the $(\prod_ks_k)^{1/K},$ and $\max_xs_k$ functions under extreme‐point decoding with uniform quantization levels $q\in\{16,32,64\}$.  Each curve corresponds to one modulation format—rectangular QAM, one‐dimensional PAM, or a hexagonal lattice—normalized to unit average power.  At high SNR, the error floors reflect the uniform‐quantizer distortion due to $\Omega_p$, while at low SNR, the gap between QAM, hexagonal, and PAM illustrates their relative robustness to AWGN.}
  \label{fig:mse-comparison}
\end{figure*}
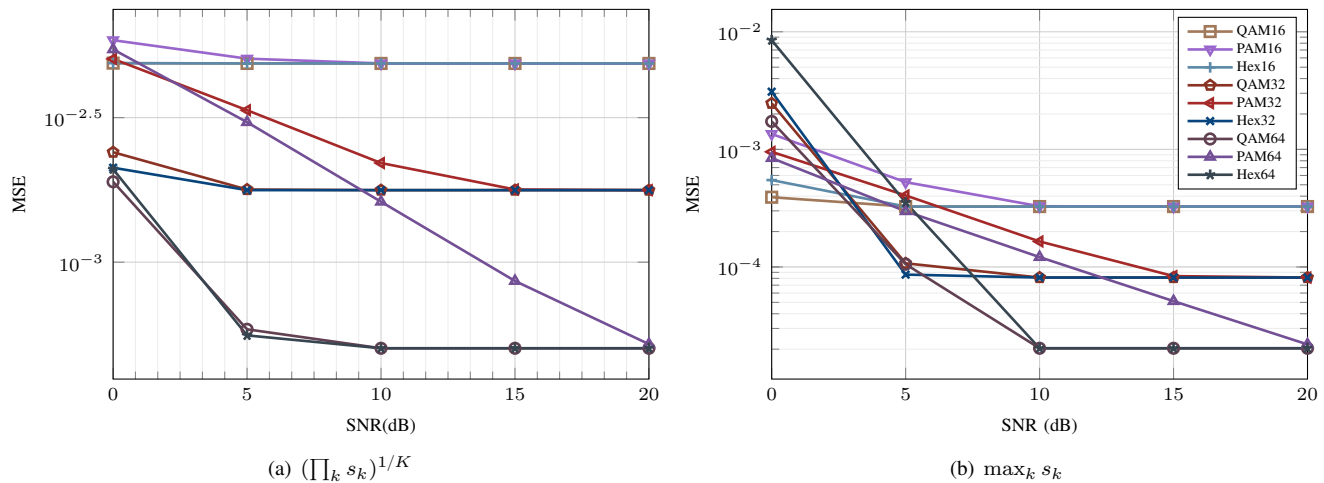

%% file: Figs/Fig_Num3.tex
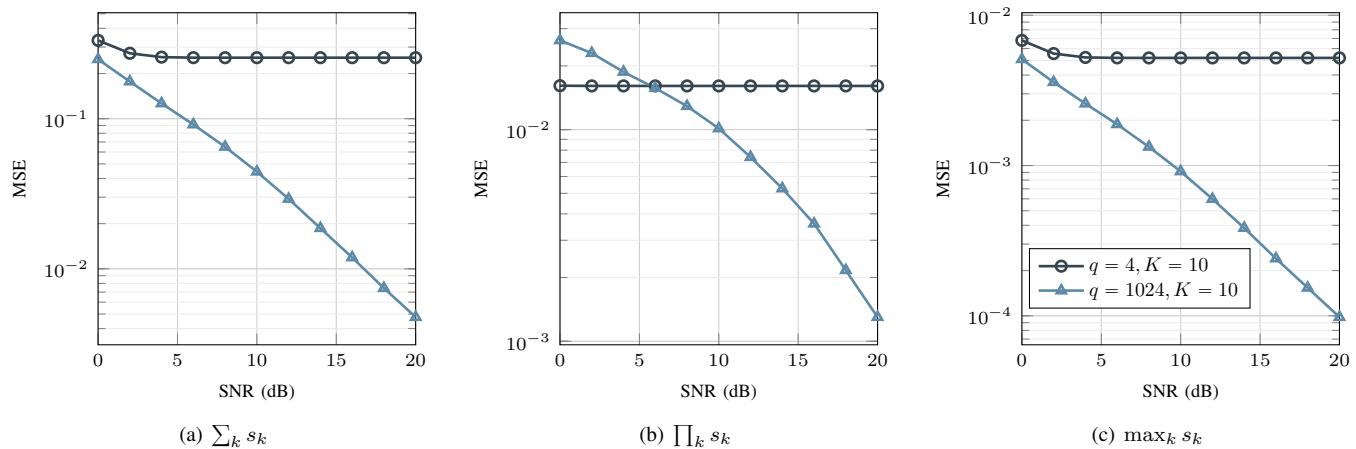
\begin{figure*}[t]
  \centering

  % --- Sum subfigure ---
  \subfigure[$\sum_k s_k$]{%
    \begin{tikzpicture}
    \begin{axis}[
      xlabel={SNR (dB)},
      ylabel={MSE},
      label style={font=\scriptsize},
      tick label style={font=\scriptsize},
      width=0.32\textwidth,
      height=6cm,
      xmin=0, xmax=20,
      ymode=log,
      legend cell align={left},
      legend style={nodes={scale=0.75,transform shape}, at={(0.98,0.5)}, anchor=east},
      ymajorgrids=true,
      xmajorgrids=true,
      grid=both,
      grid style={line width=.1pt, draw=gray!15},
      major grid style={line width=.2pt, draw=gray!40},
    ]
      \addplot[color=charcoal, line width=1pt,mark=o, mark size=2pt] table[x=snr,y=MSE_sum_s1] {Data/mse_comparison.dat}; 
        % \addlegendentry{$q=4,K=10$}
      \addplot[color=airforceblue, line width=1pt, mark=triangle, mark size=2pt]  table[x=snr,y=MSE_sum_s2] {Data/mse_comparison.dat}; 
        % \addlegendentry{$q=1024,K=10$}
    \end{axis}
    \end{tikzpicture}
  }
  \hfill
  % --- Prod subfigure ---
  \subfigure[$\prod_k s_k$]{%
    \begin{tikzpicture}
    \begin{axis}[
      xlabel={SNR (dB)},
      ylabel={MSE},
      label style={font=\scriptsize},
      tick label style={font=\scriptsize},
      width=0.32\textwidth,
      height=6cm,
      xmin=0, xmax=20,
      ymode=log,
      legend cell align={left},
      legend style={nodes={scale=0.75,transform shape}, at={(0.98,0.5)}, anchor=east},
      ymajorgrids=true,
      xmajorgrids=true,
      grid=both,
      grid style={line width=.1pt, draw=gray!15},
      major grid style={line width=.2pt, draw=gray!40},
    ]
      \addplot[color=charcoal,line width=1pt, mark=o, mark size=2pt]  table[x=snr,y=MSE_prod_s1] {Data/mse_comparison.dat}; 
        % \addlegendentry{ $q=4,K=10$}
       \addplot[color=airforceblue,line width=1pt, mark=triangle, mark size=2pt] table[x=snr,y=MSE_prod_s2] {Data/mse_comparison.dat}; 
        % \addlegendentry{ $q=1024,K=10$}
    \end{axis}
    \end{tikzpicture}
  }
  \hfill
  % --- Max subfigure ---
  \subfigure[$\max_k s_k$]{%
    \begin{tikzpicture}
    \begin{axis}[
      xlabel={SNR (dB)},
      ylabel={MSE},
      label style={font=\scriptsize},
      tick label style={font=\scriptsize},
      width=0.32\textwidth,
      height=6cm,
      xmin=0, xmax=20,
      ymode=log,
      legend cell align={left},
      legend style={nodes={scale=0.75,transform shape}, at={(0.72,0.2)}, anchor=east},
      ymajorgrids=true,
      xmajorgrids=true,
      grid=both,
      grid style={line width=.1pt, draw=gray!15},
      major grid style={line width=.2pt, draw=gray!40},
    ]
      \addplot[color=charcoal,line width=1pt, mark=o, mark size=2pt]  table[x=snr,y=MSE_max_s1] {Data/mse_comparison.dat}; 
        \addlegendentry{ $q=4,K=10$}
      \addplot[color=airforceblue,line width=1pt, mark=triangle, mark size=2pt]  table[x=snr,y=MSE_max_s2] {Data/mse_comparison.dat}; 
        \addlegendentry{ $q=1024,K=10$}
    \end{axis}
    \end{tikzpicture}
  }

  \caption{MSE versus per‐node SNR for the $\sum_k s_k,\;\prod_k s_k,$ and $\max_k s_k$ functions under two strategies. Scenario 1 employs an SDP‐designed constellation of size $q=4$ over $K=10$ users, while Scenario 2 uses an SDP‐designed constellation of size $q=1024$ (designed in the optimization Problem $\mathcal{P}_2$  for $K=2$, i.e., $p=5$ ) with $K=10$ replicas and extreme‐point decoding.}
  \label{fig:mse2-comparison}
\end{figure*}